\documentclass[11pt]{article}
\usepackage{fullpage}
\usepackage{graphicx}
\usepackage{latexsym,amsmath,amssymb,amsthm,epic,eepic,multirow}
\usepackage{natbib}

\usepackage{multirow}
\usepackage{subfigure}
\usepackage{natbib}
\usepackage{algorithm}
\usepackage{algorithmic}

\usepackage{tikz,enumerate}
\usepackage[framemethod=TikZ]{mdframed}
\usetikzlibrary{arrows}
\newdimen\arrowsize
\pgfarrowsdeclare{arcsq}{arcsq}
{
  \arrowsize=0.2pt
  \advance\arrowsize by .5\pgflinewidth
  \pgfarrowsleftextend{-4\arrowsize-.5\pgflinewidth}
  \pgfarrowsrightextend{.5\pgflinewidth}
}
{
  \arrowsize=1.5pt
  \advance\arrowsize by .5\pgflinewidth
  \pgfsetdash{}{0pt} 
  \pgfsetroundjoin   
  \pgfsetroundcap    
  \pgfpathmoveto{\pgfpoint{0\arrowsize}{0\arrowsize}}
  \pgfpatharc{-90}{-140}{4\arrowsize}
  \pgfusepathqstroke
  \pgfpathmoveto{\pgfpointorigin}
  \pgfpatharc{90}{140}{4\arrowsize}
  \pgfusepathqstroke
}

\usepackage{pgf,pgfarrows,pgfnodes,pgfautomata,pgfheaps,pgfshade}

\usepackage{hyperref}
\usepackage{listings} 
\usepackage{color} 

\definecolor{upmaroon}{rgb}{0.48, 0.07, 0.07}
\definecolor{royalazure}{rgb}{0.0, 0.22, 0.66}
\definecolor{pakistangreen}{rgb}{0.0, 0.4, 0.0}

\newcommand{\PP}{\mathbb{P}}
\newcommand{\EE}{\mathbb{E}}
\newcommand{\R}{\mathbb{R}}

\newcommand{\eps}{\varepsilon}

\newcommand{\Cov}{\mbox{Cov}}

\newcommand{\by}{\mathbf{Y}}
\newcommand{\bx}{\mathbf{X}}
\newcommand{\ba}{\mathbf{A}}
\newcommand{\argmin}{\mathrm{argmin}}
\newcommand{\pa}{\mathrm{pa}}

\newtheorem{theo}{Theorem}
\newtheorem{prop}{Proposition}

\newtheorem{corr}{Corollary}

\usepackage{color}
\newcommand\Peter[1]{{\color{red}Peter: ``#1''}}

\begin{document}

\title{Invariance, Causality and Robustness\\
  {\normalsize 2018 Neyman Lecture} \footnote{This article is covering material of
      the 2018 Neyman Lecture but also describes new developments in
      Section \ref{sec.nonlinanchor}.}}

\author{Peter B\"uhlmann \footnote{This project has received funding from
    the European Research Council under the Grant Agreement No 786461
    (CausalStats - ERC-2017-ADG).}\\
  Seminar for Statistics, ETH Z\"urich}


\maketitle


\begin{abstract}
We discuss recent work for causal inference and predictive robustness in a
unifying way. The key idea relies on a notion of probabilistic
invariance or stability: it opens up new insights for formulating causality
as a certain risk minimization problem with a corresponding notion of
robustness. The invariance itself can be estimated from general heterogeneous or
perturbation data which frequently occur with nowadays data
collection. The novel methodology is potentially useful in many
applications, offering more robustness and better ``causal-oriented''
interpretation than machine learning or estimation in standard regression
or classification frameworks. 
\end{abstract}


\section{Introduction}
Understanding the causal relationships in a system or application of
interest is perhaps the most desirable goal in terms of understanding and
interpretability. There is a rich history of developments from various
disciplines, dating back to ancient times: ``Felix, qui potuit rerum
cognoscere causas'' -- Fortunate who was able to know the causes of things
(Georgics, Virgil, 29 BC). One might think that for pure prediction tasks,
without any ambition of interpretability, knowing the causes or the causal
structure is not important. We will explain here how these problems are
related and as a consequence: (i) one can obtain ``better'' predictions when
incorporating causal aspects and (ii) one can infer causal structure from a
certain predictive perspective. 

Inferring causal structure and effects from data is a rapidly growing
area. When having access to data from fully randomized studies, Jerzy Neyman made a pioneering contribution using a potential outcome model \citep{neyman23}.
\begin{figure}[!htb]
  \begin{minipage}{0.49\textwidth }
    \begin{center}
      \includegraphics[scale = 0.4]{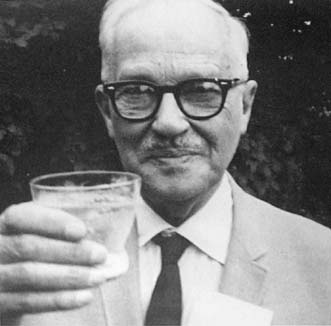}
  \end{center}
\end{minipage}
  \begin{minipage}{0.49\textwidth }
  \begin{center}
    \includegraphics[scale = 0.44]{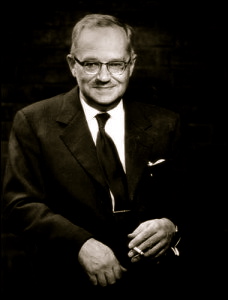}
  \end{center}
  \end{minipage}
  \caption{Jerzy Neyman (1894-1981). Besides other pioneering
    work, he has also made fundamental early contributions to causality in
    1923, in terms of mathematical formulation with the potential outcome
    model \citep{neyman23}. Left: taken from
    \url{http://www.learn-math.info/history/photos/Neyman_3.jpeg}. Right:
    taken from
    \url{https://errorstatistics.com/2017/04/16/a-spanos-jerzy-neyman-and-his-enduring-legacy-3/} 
    by A. Spanos. The photograph is hanging on the wall in the coffee room
    of the Department of Statistics at UC Berkeley.}\label{fig.neyman} 
\end{figure}

Randomized studies serve as the gold standard and the corresponding
inference of causal effects can be viewed as ``confirmatory'' due to the
fact that the underlying model assumptions are not substantially more
restrictive than for say a standard regression type problem, see for example \citet{dawid2000, pearl00, hernan2010, ruim15, vanderweele2015}. Often though, the data at hand does not come from a (fully) randomized study: the question is now whether one can still infer causal effects and under what kind of assumptions this 
is possible. A range of different approaches have been suggested, see for
example
\citet{greenland1999,robins2000,sgs00,richardson2002,hernan2006,tchetgen2012,chick02,kabu07,makapb09,hauser2015},
exhibiting different degrees of ``confirmatory'' nature for inferring
causal effects. Since causal inference is very ambitious, these techniques
should be thought as ``geared towards causality'' but not necessarily able
to infer the underlying true causal effects. Still, the point is that they
do something ``more intelligent towards causality'' than an analysis based
on a standard potentially nonlinear regression or classification
framework. We believe that this is an important area in statistics and
machine learning: in  particular, these techniques often have a more
``causal-type'' and thus more interesting interpretation than standard
machine learning methods and therefore, this topic is important in the
advent of ``interpretable machine learning''.  

\subsection{A framework based on invariance properties}
We will focus here on a particular framework with corresponding methods
which are ``geared towards'' causal solutions: with stronger assumptions
(but less strong than for some competitor methods) they infer causal
effects while under more relaxed and perhaps more realistic assumptions,
they are still providing solutions for a ``diluted form of
causality''\footnote{I am grateful to Ed George who suggested this term.}
which are often more meaningful than what is provided by regression or
classification techniques. This can be made mathematically more rigorous,
in terms of a novel form of robustness. 

The construction of methods is relies on exploiting invariance from
heterogeneous data. The heterogeneity can be unspecific
perturbations and in this sense, the current work adds to the still yet
quite small literature on statistics for perturbation data.  

\subsection{Our contribution}
The first part of the manuscript is a review of our own work in
\citet{jonipb16} and \citet{rothenhetal18} but putting the contributions
into a broader perspective. We also  
add some novel methodology on nonlinear anchor regression and present some
corresponding illustrations in Section \ref{sec.nonlinanchor}.

\section{Predicting potential outcomes, heterogeneity and worst case risk optimization}\label{sec.worstcaseoptim}

Predicting potential outcomes is a relevant problem in many application
areas.
Causality deals with a quantitative answer (a prediction) to a ``What if I do question''  or a ``What if I perturb question''.

\subsection{Two examples for prediction of potential outcomes}

The first example is from genomics \citep{steketal12}. The response variable
of interest is the flowering time of the \textit{Arabidopsis thaliana} plant
(the time 
it takes until the plant is flowering) and the 
covariates are gene expressions from 21'326 genes, that is from a
large part of the genome. The problem is to predict the flowering time of the
plant when making single gene interventions, that is, when single genes are
``knocked out''. The data is from the observational state of the system only
without any interventions. Therefore, this is a problem of predicting a
\emph{potential} outcome which has never been observed in the data. Even if
one fails to infer the true underlying causal effects when making
interventions, our viewpoint is that a prediction being better than from a
state-to-the art regression method is still very useful for
e.g. prioritizing experiments which can be done subsequently in a biology
lab, see for example \citet{mapb10} and the \citet{NMeditorial10}.

The second example is about predicting behavior of individuals when being
treated by advertisement campaigns. Such an advertisement could happen on social
media for political campaigns or various commercial products. Consider the
latter, namely an advertisement for commercial products on social
media. The response of interest is how deep an individual user clicks on the
advertisement and the subsequent web-pages, the covariates are attributes
of the user. The task is to predict the response if one would 
intervene and show to a certain user ``X'' a certain advertisement ``A'': but
there is no data for user ``X'' or similar users as ``X'' being exposed (or
treated) with advertisement ``A''. Thus, it is a problem of predicting a
\emph{potential} outcome which has never been observed in the data. As
mentioned in the genomic example above, even if we cannot infer the
underlying true causal effect of treatment with an advertisement, it is
still informative and valuable to come up with a good prediction for the
response under an intervention which
we have never seen in the data. See for example \citet{bottouetal13} or
also \citet{brodersenetal15}. 

\subsection{The heterogeneous setting with different environments}

We consider data from different \emph{observed (known)} environments, and we
sometimes refer to them also as experimental settings or sub-populations or
perturbations:
\begin{eqnarray}\label{data-setting}
(\by^e,\bx^e), e \in {\cal E}
\end{eqnarray}
with $n_e \times 1$ response vectors $\by^e$, $n_e \times p$ covariate design
matrices $\bx^e$ and $e$ denotes an environment from the space ${\cal E}$ of
observed environments. Here, $n_e$ denotes the sample size in environment
$e$. We assume that the $n_e$ samples in environment $e$ are
i.i.d. realizations of a univariate random variable $Y^e$ and a
$p$-dimensional random vector $X^e$.

\medskip\noindent
\textbf{Examples.} As a first example, consider data from 10
different countries where we know the correspondence of each data point to
one of the 10 countries. Then the space of observed (known) environments can be
encoded by the labels from ${\cal E} = \{1,2,\ldots ,10\}$. As a second
example, consider economical data which is collected over time. Different
environments or sub-populations then correspond to different blocks of
consecutive time points. Assuming that these blocks are known and given,
the space ${\cal E}$ then contains the labels for these different blocks of
sub-populations. 

\medskip
Heterogeneity can also occur outside the observed data. Thus, we consider a
space of unobserved environments
\begin{eqnarray}\label{classF}
  {\cal F} \supset {\cal E}
\end{eqnarray}
which is typically much larger than the space of observed environments ${\cal
  E}$.  

\medskip\noindent
\textbf{Examples (cont.).} In the examples from above, the space ${\cal
  F}$ could be: the 10 countries which are observed in the data and all
other countries in the world; or the sub-populations of economical
scenarios which we have observed in the data until today and all
sub-populations of future scenarios which we have not seen in the data.  

\medskip
A main task is to make predictions for new unseen environments $e
\in {\cal F}$ as discussed next. 

\subsection{A prediction problem and worst case risk
  optimization}\label{subsec.predproblem}  

We consider the following prediction problem. 
\begin{mdframed}[linewidth=0.5mm]
Predict $Y^e$ given
$X^e$ such that the prediction ``works well'' or is ``robust'' for all $e
\in {\cal F}$ based on data from much fewer environments $ e \in {\cal
  E}$. 
\end{mdframed}
Note that ${\cal F} \setminus {\cal E}$ is non-observed. The meaning of the
aim above is that one is given in the future new covariates $X^e$ from $e \in
{\cal F} \setminus {\cal E}$ and the goal is to predict the corresponding
$Y^e$. The terminology ``works well'' or is ``robust'' is understood here
in the sense of performing well in worst-case scenarios. We note that the
problem above is also related to transfer learning
\citep{pratt1993,pan2010survey,rojas2018}. 

In a linear model setting, this prediction task exhibits a relation to 
the following worst case $L_2$-risk optimization:
\begin{eqnarray}\label{worstcaserisk}
\mbox{argmin}_{b} \max_{e \in {\cal F}} \EE[|Y^e - X^e b|^2].
\end{eqnarray}
This problem has an interesting connection to causality. Before giving a
more rigorous formulation, we describe the connection in a more loose sense, for
the purpose of easier understanding. We consider the class ${\cal F}$ which
includes all heterogeneities or perturbations $e$ fulfilling two main
assumptions:
\begin{description}
\item[ad-hoc condition 1:] $e$ does not act directly on $Y^e$.
\item[ad-hoc condition 2:] $e$ does not change the mechanism between $X^e$
  and $Y^e$.
\item[ad-hoc aim:] ideally, $e$ should change the distribution of
  $X^e$.
\end{description} 
The ad-hoc conditions 1 and 2 are formulated precisely in assumption
(B$({\cal F})$) in Section \ref{subsec.invariance-causal}. Regarding the
ad-hoc aim: if there are many $e$ which change the distribution of 
$X^e$, this introduces more observed heterogeneities in ${\cal E}$ which in
term is favorable for better identification of causal effects. 

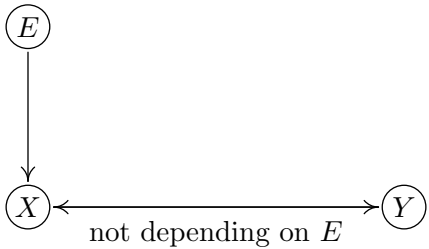
\begin{figure}[!htb]
\begin{center}
\begin{tikzpicture}[xscale=2.5, yscale=1.2, line width=0.5pt, minimum size=0.58cm, inner sep=0.3mm, shorten >=1pt, shorten <=1pt]
    \normalsize
    \draw (1,0) node(x) [circle, draw] {$X$};
    \draw (3,0) node(y) [circle, draw] {$Y$};
    \draw (1,2) node(e) [circle, draw] {$E$};
    \draw[-arcsq] (x) -- (y);
    \draw[-arcsq] (y) -- (x);
    \draw[-arcsq] (e) -- (x);
    \draw (2,-0.3) node(e) {$\mbox{not depending on $E$}$}; 
\end{tikzpicture}
\caption{Graphical illustration of the ad-hoc conditions 1, 2 and the
  ad-hoc aim. There could be also hidden confounding variables, see Section \ref{sec.anchor}.}\label{fig1}
\end{center}
\end{figure}

Figure \ref{fig1} is a graphical illustration of the ad-hoc conditions and aim
above. For this purpose, we may think that the environments $e$ are
generated from a random variable $E$. We remark that there could be also
hidden confounding variables between $X$ and $Y$: more details are given in
Section \ref{sec.anchor}.

An interesting connection to causality is then as follows:
\begin{eqnarray}\label{causal-worstcaserisk}
\mbox{argmin}_{b} \max_{e \in {\cal F}} \EE[|Y^e - X^e b|^2] =\
  \mbox{causal parameter},
\end{eqnarray}
where ${\cal F} = \{e;\ e\ \mbox{satisfies the ad-hoc conditions 1 and
  2}\}$. The definition of the causal parameter and the precise description
of the result is given later in Section \ref{subsec.invariance-causal}. The
point here is to emphasize that  
the causal parameter or the causal solution is optimizing a certain worst
case risk. This opens the door to think about causality in terms of
optimizing a certain (worst case) risk. We believe that this is a very
useful way which might ease some of the more complicated issues on
structure search for causal graphs and structural equation models. 

\section{Invariance of conditional distributions}\label{sec.invariance}

A key assumption for inferring causality from heterogeneous data as in
\eqref{data-setting} is an invariance assumption. It reads as follows: 
\begin{description}
\item[(A$({\cal E})$):] There exists a subset $S^*
  \subseteq \{1,\ldots ,p\}$ of the covariate indices (including the empty
  set) such that  
\begin{eqnarray*}
{\cal L}(Y^e|X_{S^*}^e)\ \mbox{is the same for all}\ e \in {\cal E}.
\end{eqnarray*}
That is, when conditioning on the covariates from $S^*$ (denoted by
$X^e_{S^*}$), the conditional distribution is invariant across all
environments from ${\cal E}$.
\item[(A$({\cal F})$):] Analogous but now for the much larger set of
  environments ${\cal F}$. 
\end{description}

In a linear model setting, the invariance assumption translates as
follows. There exists a subset $S^*$ and a regression coefficient vector
$\beta^*$ with $\mathrm{supp}(\beta^*) = \{j;\ \beta^*_j \neq 0\} = S^*$
such that:  
\begin{eqnarray*}
\mbox{for all $e \in {\cal E}$}:& &\ Y^e = X^e \beta^* + \eps^e,\\
& &\eps^e\ \mbox{independent of}\ X_{S^*}^e,\ \eps^e \sim F_{\eps},
\end{eqnarray*}
where $F_{\eps}$ denotes the same distribution for all $\eps^e$. That is, when
conditioning (regressing) on $X_{S^*}^e$, the resulting regression
parameter and error term distribution are the same for all environments $e
\in {\cal E}$. From a practical point of view, this is an interesting
invariance or 
stability property and the set of covariates $S^*$ plays a key component in
``stabilizing across the environments'', see also Figure
\ref{fig.screening} in Section \ref{sec.turnview}.

If the invariance assumption holds, we are sometimes interested in
describing the sets $S^*$ which fulfill invariance. We then denote by 
\begin{description}
\item[(A$_{S}({\cal E})$):] The subset $S$ fulfills invariance saying that 
\begin{eqnarray*}
{\cal L}(Y^e|X_{S}^e)\ \mbox{is the same for all}\ e \in {\cal E}.
\end{eqnarray*}
\item [(A$_{S}({\cal F})$):] analogous but now for the set of environments
  ${\cal F}$. 
\end{description}

When considering the invariance assumption for the set of unknown (future)
environments ${\cal F}$, the sets $S^*$ for which (A$_{S^*}({\cal F})$) holds
are particularly interesting as they lead to invariance and stability for new,
future environments which are not observed in the data. This is a key for
solving worst case risk optimization with respect to a class of
perturbations which can be arbitrarily strong as in
\eqref{causal-worstcaserisk}.

\subsection{Invariance and causality}\label{subsec.invariance-causal}

A main question is whether there are sets $S$ for which (A$_{S}({\cal F})$)
holds and if so, whether there are many such sets and how one can describe
them. Obviously, this depends on ${\cal F}$ and the problem then becomes
as follows: under what model ${\cal F}$ can we have an interesting
description of sets $S$ which satisfy the invariance assumption
(A$_{S}({\cal F})$).

To address this at least in part, we consider structural equation models
(SEMs):
\begin{eqnarray}\label{SEMY}
& &Y \leftarrow f_Y(X_{\pa(Y)},\eps_Y),\ \eps_Y\ \mbox{independent of}\
    X_{\pa(Y)},\nonumber\\ 
& &X \sim F_X, 
\end{eqnarray}
where $\pa(Y)$ denotes the parental set of the response variable $Y$ in
the corresponding causal influence diagram, and the distribution $F_X$ of
$X$ can be arbitrary but assuming finite second moments and positive
definite covariance matrix. 
Often in the literature, a SEM is considered for all the variables:
\begin{eqnarray}\label{SEM}
& &Y \leftarrow f_Y(X_{\pa(Y)},\eps_Y),\nonumber\\
  & &X_j \leftarrow f_j(X_{\pa(X_j)},\eps_j),
\end{eqnarray}
with $\eps_Y, \eps_1,\ldots ,\eps_p$ mutually independent. The model in
\eqref{SEM} is a special case of the SEM in \eqref{SEMY}, the former now assuming
a structural equation part for the $X$-variables. Furthermore, the
formulation in \eqref{SEMY} also allows other hidden variables which may
act on $X$ but do not have a confounding effect on $Y$. The case of hidden
confounders will be discussed later in Section \ref{sec.anchor}. 

The (direct) causal variables for $Y$ are  defined to be 
\begin{eqnarray*}
S_{\mathrm{causal}} = \pa(Y).
\end{eqnarray*}

The environments or perturbations $e$ change the distributions of $Y$ and
$X$ in model \eqref{SEMY} and we denote the corresponding random variables
by $X^e$ and $Y^e$. The ad-hoc conditions 1 and 2 from Sections
\ref{subsec.predproblem} are now formulated as follows: 
\begin{description}
  \item[(B$({\cal E})$)] The structural equation in \eqref{SEMY} remains the
  same, that is for all $e \in {\cal E}$ 
  \begin{eqnarray*}
    & &Y^e \leftarrow f_Y(X_{\pa(Y)}^e,\eps_Y^e),\ \eps_Y^e\
        \mbox{independent of}\ X_{\pa(Y)}^e,\nonumber\\ 
    & &\eps_Y^e\ \mbox{has the same distribution as}\ \eps_Y.
  \end{eqnarray*}
\item[(B$({\cal F})$)] analogous but now for the set of environments
  ${\cal F}$.  
\end{description}
We note that the distributions of $X^e$ are allowed to change. 

The following simple result describes the special role of causality with
respect to invariance.
\begin{prop}\label{prop-causeinv}
Assume a partial structural equation model as in \eqref{SEMY}. Consider
the set of environments ${\cal F}$ such that (B$({\cal F})$) holds. Then, the
set of causal variables $S_{\mathrm{causal}} = \pa(Y)$ satisfies the
invariance  assumption with respect to ${\cal F}$, that is
(A$_{S_{\mathrm{causal}}}({\cal F})$) holds. 
\end{prop}
The proof is trivial. The conditional distribution of $Y^e$ given
$X_{\pa(Y)}^e$ is given by $f_Y$ and the distribution $F_{\eps}$ of
$\eps_Y$, and these quantities do not depend on $e$.\hfill$\Box$ 

\begin{figure}
\begin{center}
  \includegraphics[scale=0.6]{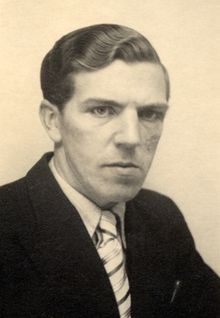}
  \caption{Trygve Haavelmo, Norwegian economist who received the Nobel
    Prize in Economic Sciences in 1989. Photo from
    \url{https://en.wikipedia.org/wiki/Trygve_Haavelmo}}\label{fig.haavelmo} 
\end{center}
\end{figure}

\medskip
In presence of hidden confounder variables, invariance and causal
structures can still be linked under certain assumptions: this will be
discussed in Section \ref{sec.anchor}. Proposition \ref{prop-causeinv} says
that causal variables lead to invariance: this has been known since a long
time, dating back to \citet{haavelmo1943}, see Figure
\ref{fig.haavelmo}. The 
result in Proposition \ref{prop-causeinv} does not say anything about other
sets of variables which satisfy the invariance assumption. 

\subsection{Invariant causal prediction}

Roughly speaking, \citet{haavelmo1943} already realized that
\begin{eqnarray*}
  \mbox{causal variables}\ \Longrightarrow\ \mbox{Invariance}.
\end{eqnarray*}
The reverse relation
\begin{eqnarray}\label{rev-impl}
  \mbox{causal structures} \Longleftarrow\ \mbox{Invariance}
\end{eqnarray}
has not been considered until recently \citep{jonipb16}. This might be due
to the fact that with nowadays large-scale data, it is much easier to infer
invariance from data and thus, the implication from invariance to causal
structures becomes much more interesting and useful. 

The problem with the reverse implication \eqref{rev-impl} is the well-known
identifiability issue in causal inference. We typically cannot identify the
causal variables $S_{\mathrm{causal}}$, unless we have very many environments
(or perturbations) or making specific assumptions on
nonlinearities \citep{hoy09,pbjoja13}, non-Gaussian distributions
\citep{shim06} or error variances \citep{petbu13}. We will address the
identifiability issue, which is often complicated in practice, in a fully
``automatic'' way as discussed next.   

The starting point is to perform a statistical test whether a subset of
covariates $S$ satisfies the invariance assumption for the observed
environments in ${\cal E}$. The null-hypothesis for testing is: 
\begin{eqnarray*}
  H_{0,S}({\cal E}):\ \mbox{assumption}\ (\mbox{A}_{S}({\cal E}))\ \mbox{holds}
\end{eqnarray*}
and the alternative is the logical complement, namely that assumption
(A$_S({\cal E})$) does not hold. It is worthwhile to point out that we only
test with respect to the environments ${\cal E}$ which are observed in the
data. To address the identifiability issue, we intersect all subsets of
covariates $S$ which lead to invariance, that is: 
\begin{eqnarray}\label{icp}
\hspace*{10mm} \hat{\cal S}({\cal E}) = \bigcap_{S} \{S;\ H_{0,S}({\cal E})\ \mbox{not rejected by test at significance level}\ \alpha\}. 
\end{eqnarray}
The specification of a particular test is discussed below in Section
\ref{subsec.tests}. The procedure in \eqref{icp} is called Invariant Causal
Prediction (ICP). The method is implemented in the \texttt{R}-package \\
\texttt{InvariantCausalPrediction} for linear models \citep{ICP-R} and
\texttt{nonlinearICP} for nonlinear models \citep{nonlinICP-R}, see also
Section \ref{subsec.tests} below.  

The computation of ICP in \eqref{icp} can be expensive. There is an
algorithm which provably computes ICP without necessarily going through all
subsets \citep{jonipb16}: in the worst-case though, this cannot be
avoided. If the dimension $p$ is large, we advocate some 
preliminary variable screening procedure based on regression for the data
pooled over all environments, see also Section \ref{subsec.tests} for the
case with linear models below. Such regression-type variable screening
procedures are valid when assuming a faithfulness condition: it ensures
that the causal variables must be a subset of the relevant regression
variables $\{j;\ X_j\ \mbox{conditionally dependent of}\
Y\ \mbox{given}\ \{X_K; k \neq j\}\}$, see for example \citet{sgs00}. 

We first highlight the property of controlling against false positive causal selections.
\begin{theo}\label{th1}\citep{jonipb16}
  Assume a structural equation model for the response $Y$ as in
  \eqref{SEMY} and that the environments or perturbations in ${\cal E}$
  satisfy the assumption (B$({\cal E})$). Furthermore, assume that the tests
  used in \eqref{icp} are valid, controlling the type I error. Then, for
  $\alpha \in (0,1)$ we have that 
  \begin{eqnarray*}
    \PP[\hat{\cal S}({\cal E}) \subseteq \pa(Y)] \ge 1- \alpha.
  \end{eqnarray*}
\end{theo} 
The interesting fact is that one does not need to care about
identifiability: it is addressed automatically in the sense that if a
variable is in $\hat{\cal S}({\cal E})$, it must be identifiable as causal
variable for $Y$, at least with controllable probability $1- \alpha$
(e.g. being equal to $0.95$). For example, even if the environments in
${\cal E}$ correspond to ineffective heterogeneities (e.g. no actual
perturbations), the statement is still valid.     

Theorem \ref{th1} does not say anything about power. The power depends on
the observed environments ${\cal E}$, besides sample size and the choice of
a test. Roughly speaking, the power increases as ${\cal E}$ becomes
larger: the more heterogeneities or perturbations, the better we can
identify causal effects and this is also true for the procedure in
\eqref{icp}. In fact, \citet[Th.2]{jonipb16} discuss cases where
the ICP method in \eqref{icp} is able to identify the all the causal
variables, i.e., where $\hat{\cal S}({\cal E}) = \pa(Y)$ asymptotically as
sample size tends 
to infinity. These special cases, where essentially all the variables are
perturbed, are far from a complete understanding of necessary and sufficient
conditions for identifiability of the causal variables. Furthermore, the
construction with the intersection in \eqref{icp} might be often
conservative. In terms of power, one wants to reject as many sets of
covariates which are violating invariance. This seems awkward at first sight
but the fact that the tests are highly dependent helps to increase the
probability that all sets $S$ which do not fulfill the invariance
hypothesis are rejected. A more quantitative statement of the latter and of
power properties for ICP in general is difficult. 

\subsubsection{Some robustness properties.}

The ICP procedure exhibits two robustness properties (and a third one
is mentioned in Section \ref{subsec.tests} below).  

\smallskip\noindent
\emph{Hidden confounding variables.} Even in presence of hidden confounding
variables (as in the scenario of Figure \ref{figIV}), we have the following:
assuming a faithfulness condition \citep[cf.]{sgs00}, 
\begin{eqnarray*} 
  \PP[\hat{S}({\cal E}) \subseteq \mathrm{an}(Y)] \ge 1- \alpha,
\end{eqnarray*}
where $\mathrm{an}(Y)$ denotes the ancestor variables of $Y$. The details
  are given in \citet[Prop.5]{jonipb16}. In practice, this is interesting as
  we would still pick up some variables which indirectly have a total
  causal effect on $Y$.

\smallskip\noindent
\emph{Direct effects of environments on $Y$.} The ad-hoc conditions 1 and
2 in Section \ref{subsec.predproblem} or the condition (B$({\cal E})$) are
violated if the environments directly affect $Y$. With a faithfulness
condition \citep[cf.]{sgs00}, we 
would then always infer that no set $S$ would fulfill the invariance
assumption and therefore, as sample size gets sufficiently large, rejecting
$H_{0,S}({\cal E})$ for all $S$, we would obtain that $\hat{\cal
  S}({\cal E}) = \emptyset$. Therefore, even under violation of the
assumption that the environments or perturbations should not act directly
on $Y$, the ICP procedure gives a conservative answer and claims no
variable to be causal. In the literature, this scenario is also known under
the name of so-called invalid instrumental variables \citep[cf.]{guo2018}.

\subsubsection{Concrete tests.}\label{subsec.tests}

We first assume that the structural equation for $Y$ in \eqref{SEMY} is linear with Gaussian error:
\begin{eqnarray*}
  Y = \sum_{j \in \pa(Y)} \beta_j X_j + \eps_Y,\ \eps_Y \sim {\cal N}(0,\sigma_Y^2)
\end{eqnarray*}
and $\eps_Y$ is independent of $X_{\pa(Y)}$. The invariance hypotheses in
$H_{0,S}({\cal E})$ then becomes:
\begin{description}
\item[$H_{0,S}({\cal E})^{\mathrm{lin-Gauss}}$:] for all $e \in {\cal E}$
  its holds that,
  \begin{eqnarray*}
    & &Y^e = X_S^e \beta_S + \eps_S^e,\ \eps_S^e\ \mbox{independent of}\
        X_S^e\ \mbox{(the same $\beta_S$ for all $e \in {\cal E}$)},\\
    & &\eps_S^e \sim F_{\eps_S}\ \mbox{(the same for all $e \in {\cal E}$)}.
  \end{eqnarray*}
\end{description}
Thanks to the Gaussian assumption, exact tests for this null-hypothesis
exist, for example with the Chow test \citep{chow1960}. This is implemented
in the \texttt{R}-package \texttt{InvariantCausalPrediction} \citep{ICP-R}.  

The variable pre-screening methods, mentioned above after the introduction
of the ICP estimator \eqref{icp}, become also much simpler in linear
models. One can use e.g. the Lasso \citep{tibs96} on the data pooled over
all observed environments and then employ the ICP estimator for all subsets
of $\hat{S}_{\mathrm{Lasso}} = \{j;\ \hat{\beta}_{\mathrm{Lasso},j} \neq
0\}$. To justify this, one needs to establish that, under
$H_{0,S}^{\mathrm{lin-Gauss}}$, $\PP[S \subseteq \hat{S}_{\mathrm{Lasso}}]
\to 1$ asymptotically: sufficient conditions for this are given in
e.g. \citet{pbvdg11}. 

When the true underlying model for the response in \eqref{SEMY} is
sufficiently nonlinear but if one uses ICP in \eqref{icp} with invariance tests
based on a mis-specified Gaussian linear model, it typically happens that
no set $S$ satisfies the invariance assumption resulting in $\hat{S}({\cal
  E}) = \emptyset$. One could use instead a testing
methodology for nonlinear and non-Gaussian models to infer whether a subset
of variables fulfills the invariance assumption. A corresponding proposal
is given in \citet{heinze2018} with the accompanying \texttt{R}-package
\texttt{nonlinICP} \citep{nonlinICP-R}.  

%

\subsubsection{Application: single gene knock-out experiments.}

We briefly summarize here the results from an application to predict single
gene interventions in yeast (\textit{Saccharomyces cerevisiae}); for
details we refer to \citet{meetal16}. The data consists of mRNA expression
measurements of $6170$ genes in yeast. We have 160 observational
measurements from wild-type yeast and 1479 interventional data arising from
single gene perturbation, where a single gene has been deleted from a
strain \citep{kemmeren2014}. The goal is to predict the expression level
of a new unseen gene perturbation, i.e., the potential outcome of a new
unseen perturbation. 

More specifically, and using the terminology of the framework outlined
before, we aim to infer some of the (direct) causal variables of a target
gene. Denote the gene expression measurements by $G_1,\ldots ,G_{6170}$. We
consider as a response variable $Y$ the expression of the $j$th gene and
the corresponding covariates $X$ the expressions of all other genes:  
\begin{eqnarray*}
& &Y = G_j,\\
  & &X = (G_1,\ldots ,G_{j-1},G_{j+1},\ldots ,G_p).
\end{eqnarray*}
The index $j \in \{1,\ldots ,6170\}$ and the covariate dimension is $p = 6169$. The aim is now to infer $\pa(Y)$, assuming a linear structural equation model as in \eqref{SEM} but now with functions $f_Y$ and $f_X$ being linear.

We construct the environments in a crude way: ${\cal E} = \{1,2\}$ where
the labels ``1'' and ``2'' denote the 160 observational and the $1479$
interventional sample points, respectively. Thus, we pool all 
the interventional samples into one environment since we have no replicates
of single gene perturbations. Other pooling schemes could be used as well:
it is typically only an issue of power how to create good environments
while the type I error control against false positive causal selections
(Theorem \ref{th1}) is still guaranteed, see also Section
\ref{subsec.envir} below. 

For validation, we do a training-test data splitting with a $K$-fold
validation scheme of the interventional data: that is, we use all
observational and $(K-1)/K$ of the interventional data with $K = 3$ or
$5$. We only consider true strong intervention effects (SIEs) where an
expression $X_k$ has a strong effect on $Y$ (the perturbed value $X_k$ are
outside of the observed data range). 

\begin{figure}[!htb]
  \begin{center}
\includegraphics[scale=0.43]{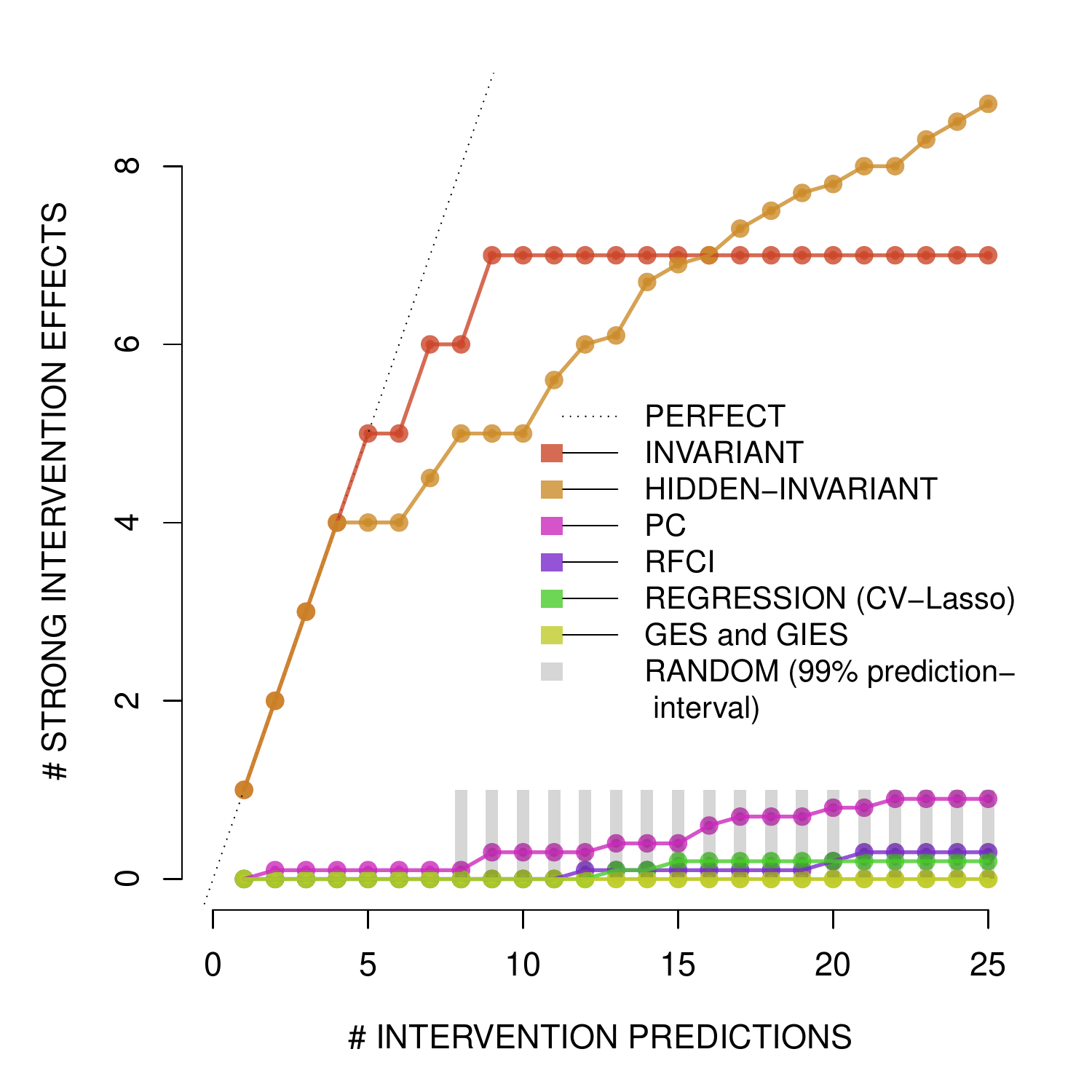}
\caption{Prediction of strong intervention effects in single gene deletion
  experiments in Saccharomyces cerevisiae (yeast). x-axis: number of predictions
  made by a method; y-axis: number of predictions being true strong
  intervention effects.. Invariant causal prediction (ICP) in red: the
  first 5 predictions are all true (and 7 among the first 9 predictions are
  true). Orange: the Causal Dantzig Selector \citep{rothenhausler2017}, an
  algorithm based on an invariance property which includes hidden
  variables. All other methods are not distinguishable from random guessing
  (gray bars). Figure is taken from \citet{meetal16}.}.\label{fig.knockout} 
\end{center}
\end{figure}

The predictions are based on ICP (with Bonferroni correction due to using
the ICP procedure many times, once for each gene being the response
variable). One then finds that 8 significant genes at corrected
significance level $\alpha = 0.05$ and 6 of them are true positive strong
intervention effects \citep{jonipb16}. What sticks out is that only very
few causal genes have been found: in a graph with 6170 nodes (corresponding
to all the genes), only 8 significant directed edges are found. When
prioritizing the most promising causal genes, Figure \ref{fig.knockout}
describes ROC-type curves: here ICP is supplemented with stability
selection \citep{mebu10} on top of it for creating a ``stabilized''
ranking. 

\subsubsection{Unknown environments.}\label{subsec.envir}

If the environments $e \in {\cal E}$ are not known, one can try to estimate
them from data. The type I error control against false positive causal
selections holds as long as the estimated partition $\hat{\cal E}$ does not
involve descendant variables of the response $Y$: for example, one could
use some clustering algorithm based on non-descendants of $Y$. 

In practice, it is sometimes reasonable to assume that certain variables
are non-descendants of $Y$. A canonical case is with time-sequential data:
then, the environments can be estimated as different blocks of data at
consecutive time points: this is some kind of a change point problem but
now aimed for most powerful discovery with ICP. The methodology with
time-sequential data is developed and analyzed in \citet{pfister2018} and
implemented in the \texttt{R}-package \texttt{seqICP} \citep{seqICP-R}.   

\section{Anchor regression: relaxing conditions}\label{sec.anchor}

The main concern with ICP in \eqref{icp} and the underlying invariance
principle is the violation of the assumption in (B$({\cal E})$),
and thus also of the ad-conditions 1 and 2 from Section
\ref{subsec.predproblem}. Such a violation can happen under various
scenarios and we mention a few in the following.

It could happen that only approximate instead of exact invariance
holds. This would imply that we should only search for approximate
invariance, something which we will incorporate in the anchor regression
methodology described below in Section \ref{subsec.anchorest}. Another
scenario, say in a linear model, is that invariance occurs in the
null-hypothesis $H_{0,S}({\cal E})^{\mathrm{lin-Gauss}}$ for the
parameter $\beta_S^e \equiv \beta_S$ but with residual
distributions $\eps_S^e$ which change for varying $e$; or vice-versa with
invariant residual distribution but different regression parameters for
varying $e$. This could be addressed in ICP by testing only either the
parameter- or residual-part, where invariance is assumed to hold for the causal
variables. 

\begin{figure}[!htb]
\begin{center}
\begin{tikzpicture}[xscale=2.5, yscale=1.2, line width=0.5pt, minimum size=0.58cm, inner sep=0.3mm, shorten >=1pt, shorten <=1pt]
    \normalsize
    \draw (1,0) node(x) [circle, draw] {$X$};
    \draw (3,0) node(y) [circle, draw] {$Y$};
    \draw (0.5,2) node(e) [circle, draw] {$E$};
    \draw (2,2) node(h) [circle, draw] {$H$};
    \draw[-arcsq] (x) -- (y);
    \draw[-arcsq] (y) -- (x);
    \draw[-arcsq] (e) -- (x);
    \draw[-arcsq] (h) -- (x);
    \draw[-arcsq] (h) -- (y);
\end{tikzpicture}
\caption{Graphical illustration with hidden confounding variables $H$. It
  corresponds to the instrumental variables regression model, where the
  instruments are now the environments.}\label{figIV}   
\end{center}
\end{figure}
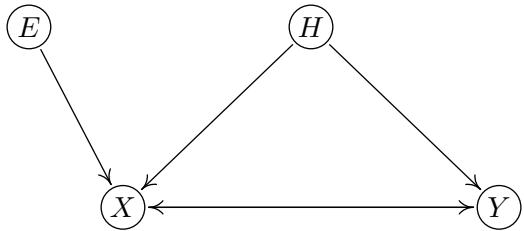

Perhaps the most prominent violation is in terms of hidden confounding
variables $H$. The influence diagram from Figure \ref{fig1} can then be
extended to the situation of an instrumental variables (IV) regression
model \citep{bowdenturk90,angristetal96,ruim15}, illustrated in Figure
\ref{figIV}.
Now, it is more convenient 
to think of the environments (or instruments) as random variables and we
also model all random variables in the system in terms of a structural
equation model (unlike as in \eqref{SEMY}, where we have only one
structural equation for $Y$): instead of \eqref{SEM}, we consider now the
IV regression model
\begin{eqnarray*}
& &Y \leftarrow f_Y(X_{\pa_X(Y)},H,\eps_Y),\nonumber\\
  & &X_j \leftarrow f_j(X_{\pa_X(X_j)},H,E,\eps_j),
\end{eqnarray*}
where $H, E, \eps_Y, \eps_1,\ldots, \eps_p$ are mutually independent of
each other. The variable $H \in \R^r$ is hidden (not observed) and possibly
confounding between $X$ and $Y$ (if some components of $H$ are descendants
of $X$ or $Y$, these are not relevant for inferring the effect from $X$ to
$Y$ and hence w.l.o.g. $H$ is a source node).   
Here, $\pa_X(\bullet)$ denotes the parental variables
$X$-variables (variables which are parents of $\bullet$ and from the set
$\{X_1,\ldots ,X_p\}$).  
The main assumption in the IV regression model requires that the instruments or
environments do not directly influence the response variable $Y$ nor the
hidden confounders $H$ (this is an extension of the ad-hoc conditions 1 and
2 from Section \ref{subsec.predproblem}); and ideally, they would influence
or change the $X$ variables in a sufficiently strong way (as with the
ad-hoc aim in Section \ref{subsec.predproblem}). We are not going into more
details from the vast literature on IV models with e.g. weak instruments, invalid
instruments or partially identifiable parameters, see for example
\cite{stock2002,murray2006,kang2016,guo2018}. Instead, we will relax this
main assumption for an instrument as discussed next.    

\subsection{The anchor regression model}

We will allow now that the environments can act directly also on $H$ and
$Y$, relaxing a main assumption in IV regression models. In the terminology
of IV regression, we thus consider the case with so-called invalid
instruments \citep[cf.]{guo2018}. This is an ill-posed situation for causal
inference (from $X$ to $Y$), yet it is still possible to obtain more 
meaningful results than what is obtained from standard regression
methodology, see Section \ref{subsec.diluted-causality}. 

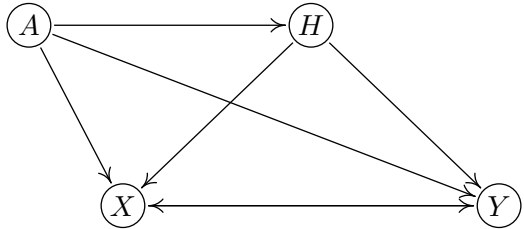
\begin{figure}[!htb]
  \begin{center}
\begin{tikzpicture}[xscale=2.5, yscale=1.2, line width=0.5pt, minimum size=0.58cm, inner sep=0.3mm, shorten >=1pt, shorten <=1pt]
    \normalsize
    \draw (1,0) node(x) [circle, draw] {$X$};
    \draw (3,0) node(y) [circle, draw] {$Y$};
    \draw (0.5,2) node(a) [circle, draw] {$A$};
    \draw (2,2) node(h) [circle, draw] {$H$};
    \draw[-arcsq] (x) -- (y);
    \draw[-arcsq] (y) -- (x);
    \draw[-arcsq] (a) -- (x);
    \draw[-arcsq] (h) -- (x);
    \draw[-arcsq] (h) -- (y);
    \draw[-arcsq] (a) -- (y);
    \draw[-arcsq] (a) -- (h);
\end{tikzpicture}
\caption{Graphical structure of the anchor regression model in \eqref{mod.anchor}, where $A$ denotes the anchors which have been referred to as environments before. Note that the anchor variables are source nodes in the graph.}\label{fig.anchor}
\end{center}
\end{figure}

Instead of using the terminology ``environment'' we now us the word
``anchor'' (or anchor variable), for reasons which become more clear 
below. The structure of an anchor regression model is given by the graph in
Figure \ref{fig.anchor}.
The anchor regression model, for simplicity
here only in linear form, is defined as follows: it is a structural
equation model of the form,
\begin{eqnarray}\label{mod.anchor}
\begin{pmatrix}
           X \\
           Y \\
           H
         \end{pmatrix} = B \begin{pmatrix}
           X \\
           Y \\
           H
         \end{pmatrix} + \eps + M A
\end{eqnarray}
We assume that all the variables are centered with mean zero. 
There could be feedback cycles and the graph of the structure could have
cycles. In the latter case, we assume that $I - B$ is invertible (which
always holds if the graph is acyclic). The main assumption is that $A$ is a
source node and thus, the contribution of $A$ enters as an additional
linear term $MA$. Because of this, we use the terminology ``anchor'': it is
the anchor which is not influenced by other variables in the system and
thus, it remains as the ``static pole''. 

As mentioned above, one cannot identify the causal parameter $B_{Y,X}$, the
row and columns corresponding to $Y$ and $X$ in the matrix $B$. However, we
we will discuss in Section \ref{subsec.robustness}, that we can still get an
interesting solution which 
optimizes a worst case risk over a class of scenarios or
perturbations ${\cal F}$, using the terminology and spirit of
\eqref{worstcaserisk}. 

\subsection{Causal regularization and the anchor regression
  estimator}\label{subsec.anchorest} 

We are particularly interested in the structural equation for $Y$ in the model \eqref{mod.anchor} which we write as
\begin{eqnarray*}
  Y = X^T \beta + H^T \alpha + A^T \xi + \eps_Y,
\end{eqnarray*}
with $X \in \R^p,\ H \in \R^q$ and $A \in \R^r$. 

In the instrumental variables regression model where $A$ would directly
influence $X$ only, it holds that $H, \eps_Y, A$ are mutually independent (but
not so in the anchor regression model) and this then implies that 
\begin{eqnarray*}
  Y - X^T \beta\ \mbox{is uncorrelated of}\ A.
\end{eqnarray*}
Actually, we could substitute uncorrelatedness with independence in the IV
model.

In cases where the causal parameter is non-identifiable, one could look for
the solution 
\begin{eqnarray}\label{L2-uncor}
  \mbox{argmin}_b \EE[(Y - X^T b)^2]\ \mbox{such that}\ \mbox{Corr}(A,Y -
  X^Tb) = 0. 
\end{eqnarray}
This leads to a unique parameter (assuming that $\Cov(X)$ is positive
definite), and there is a simple pragmatic principle behind it. This
principle and the property of uncorrelatedness of the residual term with
the anchor variables $A$ also plays a key role in the anchor regression
model for a class of so-called shift perturbations.

Similar to the idea in \eqref{L2-uncor}, we define the anchor regression
estimator by using a regularization term, referred to as causal
regularization, which encourages orthogonality or uncorrelatedness of the
residuals with the anchor variables $A$. We denote the data quantities by
the $n \times 1$ response vector $\by$, the $n \times p$ covariate design
matrix $\bx$ and the $n \times q$ matrix $\ba$ of the observed
anchor variables. Let $\Pi_{\ba}$ the projection
in $\R^n$ onto the column space of $\ba$. In practice, if the columns of
$\bx$ and $\by$ are not centered, we would include an intercept column in
$\ba$. We then define  
\begin{eqnarray}\label{anchor-est}
  & \hat{\beta}(\gamma) = \mbox{argmin}_b \left(\|(I - \Pi_{\ba})(\by - \bx
  b)\|_2^2/n + \gamma \|\Pi_{\ba}(\by - \bx b)\|_2^2/n \right). 
\end{eqnarray} 
We implicitly assume here that $\mathrm{rank}(\ba) = r < n$. For $\gamma =
1$, $\hat{\beta}(1)$ equals the ordinary least squares estimator, for
$\gamma \to \infty$ we obtain the two-stage least squares procedure from IV
regression and for $\gamma \to 0$ we adjust for the anchor variables in
$A$. The properties of the anchor regression estimator in
\eqref{anchor-est} are discussed next and make the role of the tuning
parameter more clear.

The criterion function on the right-hand side of \eqref{anchor-est} is a
convex function in $b$; for high-dimensional scenarios, we can add an
$\ell_1$-norm penalty, or any other sparsity inducing penalty:
\begin{eqnarray}\label{anchor-estLasso}
& &\ \hat{\beta}(\gamma) = \mbox{argmin}_b \left(\|(I - \Pi_{\ba})(\by - \bx
    b)\|_2^2/n + \gamma \|\Pi_{\ba}(\by - \bx b)\|_2^2/n + \lambda \|b\|_1 \right).
\end{eqnarray}

The computation of the anchor regression estimator is trivial. We simply
transform the variables $Y$ and $X$,
\begin{eqnarray*}
  & &\tilde{Y} = W_{\gamma} Y,\ \tilde{X} = W_{\gamma}X,\\
  & &W_{\gamma} = I - (1 - \sqrt{\gamma}) \Pi_{\ba}.
\end{eqnarray*}
The anchor regression estimator in \eqref{anchor-est} or
\eqref{anchor-estLasso} is then given by ordinary least squares or Lasso
for the regression of $\tilde{Y}$ versus $\tilde{X}$.

\subsection{Shift perturbations and robustness of the anchor regression estimator}\label{subsec.robustness}   

The anchor regression estimator solves a worst case risk optimization
problem over a class of shift perturbations. 

We define the system under shift perturbations $v$ by the same equations as
in \eqref{mod.anchor} but replacing the term $M A$ from the contributions of
the anchor variables by a deterministic or stochastic perturbation vector
$v$. That is, the system under shift perturbations satisfies: 
\begin{eqnarray*}
\begin{pmatrix}
           X^v \\
           Y^v \\
           H^v
         \end{pmatrix} = B \begin{pmatrix}
           X^v \\
           Y^v \\
           H^v
         \end{pmatrix} + \eps + v = (I-B)^{-1} (\eps + v). 
\end{eqnarray*}
The shift vector $v$ is assumed to be in the span of $M$, that is $v = M
\delta$ for some vector $\delta$.
The class of considered perturbations, denoted earlier as ${\cal F} \in
\eqref{classF}$, are shift perturbations as follows:
\begin{eqnarray}\label{shift-pertclass}
  C_{\gamma} = \{v;& &\hspace*{-4mm}v = M \delta\ \mbox{for random or
                       deterministic}\ 
  \delta,\ \mbox{uncorrelated with}\ \eps\nonumber\\
  & &\hspace*{-4mm}\mbox{and}\ \EE[\delta \delta^T]
  \preceq \gamma \EE[A A^T]\}.
\end{eqnarray}
Thus, $C_{\gamma}$ contains shift perturbations whose length $\|v\|_2^2$ is
typically $O(\gamma)$ as $\gamma \to \infty$. 

For the case with $\gamma \to \infty$ one can characterize shift-invariance
of residuals as follows.
\begin{prop}\citep[Th.3]{rothenhetal18}\label{prop.shiftinv}
  Assume that $\EE[A A^T]$ is positive definite. Consider
  \begin{eqnarray*}
    I = \{b \in \R^p;\ \EE[A (Y - X^Tb)] = 0\}.
  \end{eqnarray*}
  Since $Y$ and $X$ have mean zero it follows also that $\EE[Y - Xb] = 0$
  and hence $\EE[A (Y - X^Tb)] = \mbox{Corr}(A,Y - Xb)$. Then, 
  \begin{eqnarray*}
      b \in I \Longleftrightarrow\ Y^v - (X^v)^T b\ \mbox{has the same
    distribution for all}\ v \in \mathrm{span}(M).
  \end{eqnarray*}
\end{prop}

With the goal to make the residuals invariant (for the class of shift
perturbations), we aim to estimate the regression parameter $\beta$ such
that the residuals are encouraged to be fairly uncorrelated with $A$. This
leads to the construction of the anchor regression estimator in
\eqref{anchor-est} or \eqref{anchor-estLasso}.  

A more general result than Proposition \ref{prop.shiftinv} is possible,
uncovering a robustness property of the anchor regression estimator. 
We focus first on the population case. We denote by $P_A(\cdot)$
the projection operator, namely $P_A(Z)= \EE[Z|A]$. In the anchor
regression model \eqref{mod.anchor} with $(I-B)$ being invertible, $P_A(Y)$
and $P_A(X)$ are linear functions in $A$. The population version of the
anchor regression estimator is 
\begin{eqnarray}\label{anchor-pop}
& &\ \beta(\gamma) = \mbox{argmin}_{b} \left( \EE[((I - P_A)(Y - X^T
  b))^2] + \gamma \EE[(P_A(Y - X^T b))^2] \right). 
\end{eqnarray}
Then, the following fundamental result holds.
\begin{theo}\citep[Th.1]{rothenhetal18}\label{th.anchorrob}
  For any $b \in \R^p$ it holds that
  \begin{eqnarray*}
    \ \ \sup_{v \in C_{\gamma}} \EE[(Y^v - (X^v)^T b)^2] = \EE[((I - P_A)(Y -
    X^T b))^2] + \gamma \EE[(P_A(Y - X^T b))^2].
    \end{eqnarray*}
  \end{theo}
Thus, Theorem \ref{th.anchorrob} establishes an exact duality between the
causal regularized risk (which is the population version of the objective
function for the estimator in \eqref{anchor-est}) and worst case risk over
the class of shift perturbations. The regularization parameter equals the
``strength'' of the shift, as defined in \eqref{shift-pertclass}: regarding
its choice, see Section \ref{subsec.choogamma}. A useful interpretation of
the theorem is as follows. The worst case risk over shift perturbations can
be considered 
as the one corresponding to future unseen data: this risk for future unseen
data can be represented as a regularized risk for the data which we observe
in the training sample. 
We further note that Theorem \ref{th.anchorrob} holds for any $b$, and thus
it also holds when taking the ``argmin'' on both sides of the equation. We then
obtain that the population version $\beta(\gamma)$ in
\eqref{anchor-pop} is the minimizer of the worst case risk:
\begin{eqnarray*}
  \beta(\gamma) = \mbox{argmin}_{b} \sup_{v \in C_{\gamma}} \EE[(Y^v -
  (X^v)^T b)^2].
\end{eqnarray*}

One can argue that also in the finite sample high-dimensional sparse scenario,
the anchor regression estimator in \eqref{anchor-est} or
\eqref{anchor-estLasso} are asymptotically
optimizing the worst case risk:
\begin{eqnarray*}
  \sup_{v \in C_{\gamma}} \EE[(Y^v - (X^v)^T \hat{\beta}(\gamma))^2] \le
  \min_b \sup_{v \in C_{\gamma}} \EE[(Y^v - (X^v)^T b)^2] + \Delta,
\end{eqnarray*}
where, under suitable conditions, $\Delta \to 0$ as $n \to \infty$, or $p
\ge n \to \infty$ in the high-dimensional scenario. The details are
given in \citet[Sec.4]{rothenhetal18}. 

\subsubsection{Choosing the amount of causal regularization.}\label{subsec.choogamma}
The value of $\gamma$ in the estimator \eqref{anchor-est} or
  \eqref{anchor-estLasso} relates to the class of shift perturbations over
  which we achieve the best protection against the worst case, see Theorem
  \ref{th.anchorrob}. Thus, we could decide a-priori how much protection we
  wish to have or how much perturbation we expect to have in new test
  data.

Alternatively, we could do some sort of cross-validation. If the anchor
variable encodes discrete environments, we could leave out data from one or
several environments and predict on the left-out test-data optimizing the
worst case error. If the anchor
variables are continuous, the following characterization is useful. 

The value $\gamma$ in the causal regularization has also an interpretation
as a quantile. Assuming a joint Gaussian distribution of the variables $Y, X$ and $A$ in the model \eqref{mod.anchor}, it holds that
\begin{eqnarray}\label{insamp-quant}
  & &\alpha-\mbox{quantile of}\ \EE[(Y - X^T b)^2|A]\nonumber\\
  &=& \EE[((I - P_A)(Y -
  X^T \beta))^2] + \gamma \EE[(P_A(Y - X^T \beta))^2],\nonumber\\
& &\mbox{for}\ \gamma = \alpha-\mbox{quantile of}\ \chi^2_1.
\end{eqnarray}
The right-hand side also equals a worst case risk over shift perturbations, as stated in Theorem \ref{th.anchorrob}. Therefore, the relation above links the in-sample (for the non-perturbed data) quantiles to the out-sample (for the perturbed data) worst case risk.   
The exact correspondence of $\alpha$ and $\gamma$ might not hold for more
general situations. However, the qualitative correspondence is that a large
$\alpha$ (high quantile) corresponds to a high value $\gamma$ for the
regularization term. Thus, one could choose a quantile value $\alpha$,
e.g. $\alpha = 0.95$, for
the quantile of the conditional expectation of the squared error $\EE[(Y -
X^T b)^2|A]$ and then calculate the $\gamma$ which optimizes this
quantile. Under a Gaussian assumption, we can estimate the quantities
replacing expectations by mean squared test samples. This result also
indicates that anchor regression with a large value of $\gamma$ should
result in good values for the high quantile of the squared prediction error
(unconditional on $A$).

\subsubsection{Diluted form of causality.}\label{subsec.diluted-causality}
In the anchor regression model in general, it is impossible to infer the
direct causal parameter $\beta$ from $X$ to $Y$. If the assumptions for
instrumental variables regression are fulfilled, i.e., no direct effects
from $A$ to $H$ and to $Y$ and $\mathrm{rank}(A) \ge \mathrm{dim}(X) = p$,
then the anchor regression estimator with $\gamma \to \infty$ equals the
unique two stage least squares estimator and consistently infers $\beta$;
in particular, we also have that $\beta(\gamma \to \infty) = \beta$. 

If the IV assumptions do not hold, for example in presence of invalid
instruments where the anchor variables directly affect $Y$ or $H$, the
parameter $\beta(\gamma)$ with $\gamma \to \infty$ or $\gamma$ being large is
still a much more meaningful quantity than the standard regression
parameter (with $\gamma = 1$). For large values of $\gamma$, the
corresponding $\beta(\gamma)$ is minimizing a worst case risk over a class
of large shift perturbations. This parameter and its entries with large
values corresponding to important variables is interesting in many
applications: the variables (with corresponding large parameter components
of $\hat{\beta}(\gamma)$) are ``key drivers'' in a system of interest to
explain the response $Y$ in a stable 
manner over many perturbations. In fact, for $\gamma \to \infty$, we define 
\begin{eqnarray*}
  \mathrm{supp}(\beta(\gamma \to \infty))
\end{eqnarray*}
to be the set of variables which are ``diluted causal'' for the response
$Y$ (the variables which are relevant for $Y$ in the framework of ``diluted
causality''). 

\subsection{Some empirical illustrations}

\begin{figure}[!htb]
\begin{center}
\includegraphics[scale=0.7]{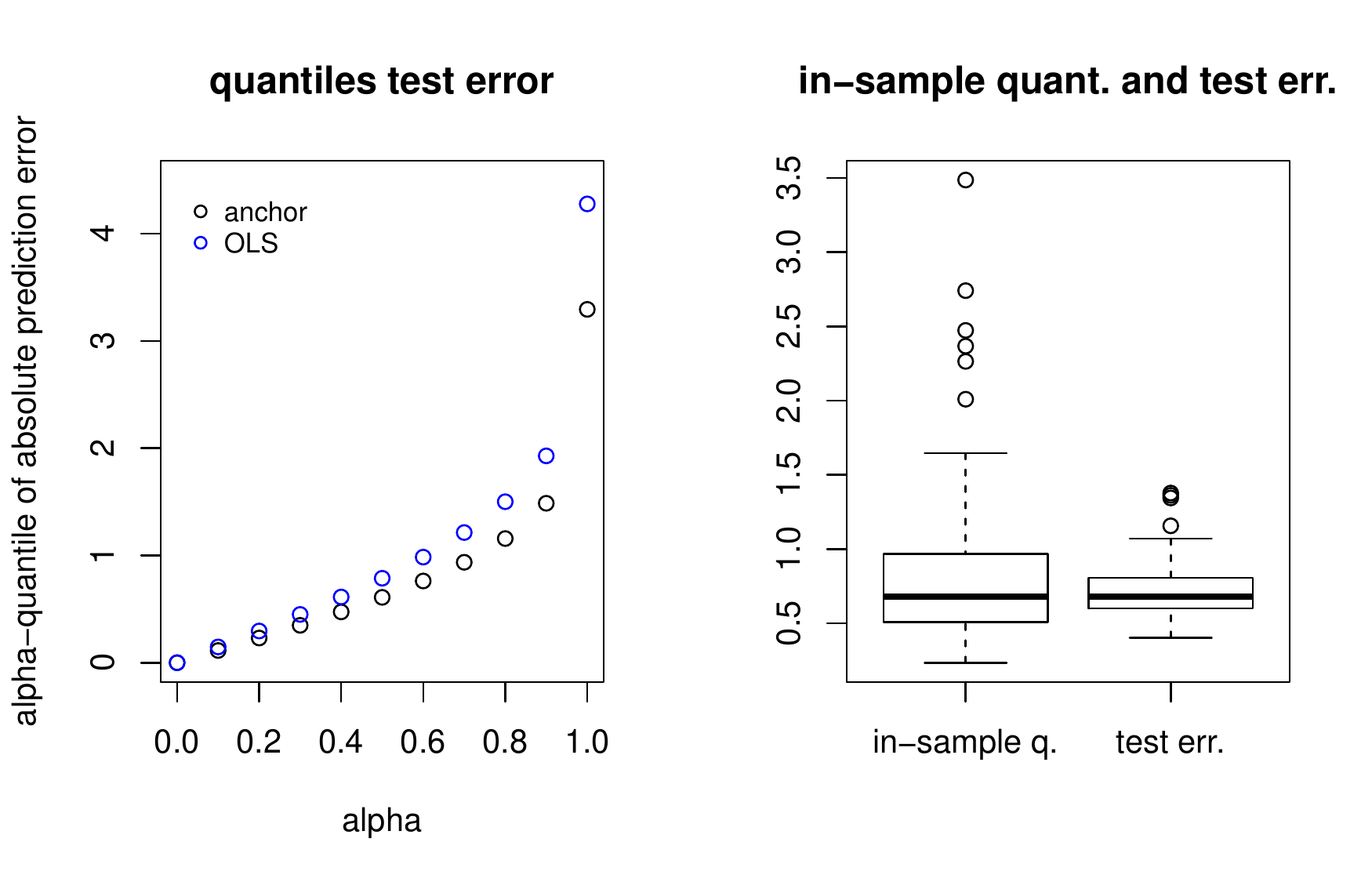}
\caption{Left: empirical $\alpha$-quantiles of $|Y_{\mathrm{out},i} -
  \hat{Y}_{\mathrm{out},i}|$ for $i = 1,\ldots, n_{\mathrm{out}} = 2000$,
  averaged over 100 independent simulation runs. Right: estimated in-sample
  $\alpha$-quantile of $\EE[(Y - X^T \hat{\beta}(\gamma))^2|A]$ in
  \eqref{insamp-quant} with $\alpha = 0.9918$ corresponding to $\gamma = 7$
  (left boxplot) and out-sample (with perturbation) mean squared prediction
  error (right boxplot), for the 100 independent simulation
  runs. }\label{fig.linanchor}   
\end{center}
\end{figure}

We illustrate the performance and behavior of the estimator \eqref{anchor-est} in the linear anchor regression model \eqref{mod.anchor}. We consider the case where the anchors are invalid instruments and low-dimensional and hence, inferring the causal effects from $X$ to $Y$ is impossible.  The model for the variables $A, H$ and $X$ is as in model (M3) described later in Section \ref{subsec.varimp}, with $\mathrm{dim}(A) = r = 2$, $\mathrm{dim}(H) = q = 1$ and $\mathrm{dim}(X) = p = 10$. The structural equation for the response is
\begin{eqnarray*}
  Y \leftarrow 3X_2 + 3X_3 + H - 2A_1 + \eps_Y,\ \eps_Y \sim {\cal N}(0,0.25^2).
\end{eqnarray*}
The training sample size is chosen as $n=300$. The test sample is constructed with the same structure but now the anchor variables $A$ are multiplied by the factor $\sqrt{10}$ and the test sample size is chosen as $n_{\mathrm{out}} = 2000$. It is instructive to describe here how the anchor variables $A$ act on $X$: the model is
\begin{eqnarray*}
  X_{j} \leftarrow A_1 \gamma_1 + A_2 \gamma_2 + H + \eps_{X_j},\ \eps_{X_j} \sim {\cal N}(0,1),
\end{eqnarray*}
where $\gamma_1, \gamma_2$ are coefficients which have been sampled
i.i.d. from ${\cal N}(0,1)$. The equation above changes in the test sample
where we multiply $A_1$ and $A_2$ with the factor $\sqrt{10}$ which results
in perturbations for the $X_j$ variables. Figure \ref{fig.linanchor}
describes the quantiles of the absolute out-sample prediction error
$|Y_{\mathrm{out},i} - X_{\mathrm{out},i}^T \hat{\beta}(\gamma)|$ where
$\gamma = 7$ has been pre-specified. We also show the empirical relation
between the in-sample quantile in \eqref{insamp-quant} and the out-sample
mean squared prediction error $\frac{1}{2000} \sum_{i=1}^{2000}
(Y_{\mathrm{out},i} - X_{\mathrm{out},i}^T \hat{\beta}(\gamma))^2$.  
We conclude from the left panel of Figure \ref{fig.linanchor} that the
anchor regression estimator exhibits a substantially better prediction
performance under perturbation out-sample scenarios than the ordinary least
squares estimator. If the out-sample data would be generated as the
in-sample training data, that is without new perturbations in the test
data, there would be no gain, or actually a slight loss, of anchor
regression over OLS (empirical results not shown here). Anchor regression
only pays-off for prediction if some perturbations happen in new future
data points which amplify the effect of heterogeneity (generated from the
anchor variables $A$) in the future test data. This is briefly discussed
next.

\subsection{Distributional robustness}

Anchor regression and causality can be viewed from the angle of distributional robustness \citep{heinze2017,meinshausen2018}. Distributional robustness refers to optimizing a worst case risk over a class of distributions:
\begin{eqnarray*}
\argmin_{\theta} \max_{P \in {\cal P}} \EE[\ell(Z;\theta)],
\end{eqnarray*}
where $\ell(\cdot;\cdot)$ denotes a loss function, $Z$ is the random variable generating a data point (e.g. $Z = (Y,X)$), $\theta$ is an unknown (potentially high- or infinite-dimensional) parameter and ${\cal P}$ is a class of probability distributions.

A typical choice for the class of distributions is
\begin{eqnarray*}
  {\cal P} = \{P;\ d(P,P_0) \le \rho\},
\end{eqnarray*}
where $P_0$ is the reference distribution, for example being the empirical measure, $d(\cdot,\cdot)$ is a metric, for example the Wasserstein distance, and $\rho$ is a pre-specified radius, see for example \citet{sinha2017,gao2017}.

For causality and anchor regression, the class of distributions ${\cal P}$
is given by a causal or ``anchor-type'' model consisting of perturbation
distributions. Theorem \ref{th.anchorrob} describes the
connection more explicitly: the class ${\cal P}$ consists of amplifications
of the observed heterogeneity in the data. This, because the perturbation
distributions arise from shifts $v \in \mathrm{span}(M)$, thus being shifts
in the direction of the effects from the observed anchor variable
contribution which equals the term $MA$ in the model \eqref{mod.anchor};
and the strength of the shifts in the perturbations is given by the
parameter $\gamma$ which has an analogous role as the radius $\rho$ in the
definition of ${\cal P}$ above. Thus, with anchor regression, the class of
distributions is not pre-defined via a metric $d(\cdot,\cdot)$ and a radius
$\rho$ but rather through the observed heterogeneities in
$\mathrm{span}(M)$ and a strength of perturbations or ``radius'' $\gamma$.   

\section{Nonlinear anchor regression}\label{sec.nonlinanchor}

We present here a methodology for anchor regression generalizing the
linear case. A core 
motivation is to design an algorithm for which any ``machine learning''
technology for regression can be plugged-in, including for example Random
Forests or even Deep Neural Nets. We will argue that in presence of
heterogeneity, essentially ``any'' of these machine learning methods can
be improved using additional causal regularization. 

We consider a nonlinear anchor regression structural equation model where
the dependence of $Y$ on $X$ is a nonlinear function: 
\begin{eqnarray}\label{nonlin.anchor.model}
& &X \leftarrow M_{X} A + B_{X,H} H + \eps_X,\nonumber\\
& &Y \leftarrow f(X) + M_{Y} A + B_{Y,H} H + \eps_Y,\nonumber\\
& &H \leftarrow M_H A + \eps_H.
\end{eqnarray}
We can consider more general functions although a too high
degree of nonlinearity can become 
more difficult with the anchor regression algorithm presented below. This
is discussed after Corollary \ref{corr1}. We note that with $M_{Y}$
and $M_H$ being equal to zero we have a nonlinear instrumental variables
regression model with linear dependences on the instruments (or the
anchors) $A$ and the hidden variables $H$. 

\subsection{The objective function and the algorithm} 

Consider a nonlinear regression function $f$, defined as 
\begin{eqnarray*}
f(x) = \EE[Y|X=x]
\end{eqnarray*}
which is a map from ${\cal X}$ to ${\cal Y}$, where ${\cal
  X}$ and ${\cal Y}$ are the domains of $X$ and $Y$, respectively. Given
data $(Y^{(1)},X^{(1)}),\ldots ,(Y^{(n)},X^{(n)})$, many estimation methods
or algorithms for $f$ can be written in the form  
\begin{eqnarray}\label{nonlin-est}
\hat{f} = \argmin_{f \in C} \frac{1}{2n} \sum_{i=1}^n (Y^{(i)} -
  f(X^{(i)}))^2 = \argmin_{f \in C} \frac{1}{2n}\|Y - f\|_2^2
\end{eqnarray}
where $C$ denotes a suitable sub-class which incorporates certain restrictions
such as smoothness or sparsity. On the right-hand side we have used a slight
abuse of notation where $f = (f(X^{(1)}),\ldots ,f(X^{(n)}))^T$ denotes the vector
of function values at the observed $X^{(1)},\ldots ,X^{(n)}$. As will be seen
below, the estimation algorithm is not necessarily of the form as in
\eqref{nonlin-est} but we use this formulation for the sake of simplicity. 

Using the abbreviated notation with $f$ mentioned above, the
nonlinear anchor regression estimator is defined as: 
\begin{eqnarray}\label{estimator}
& &\hat{f}_{\mathrm{anchor}} = \argmin_{f \in C} G(f),\nonumber\\
& &G(f) = G_{\gamma}(f) = \frac{1}{2} \left(\|(I - \Pi_{\ba})(Y - f)\|_2^2/n
  + \gamma \|\Pi_{\ba}(Y - f)\|_2^2/n \right).
\end{eqnarray}
As in \eqref{anchor-est}, $\Pi_{\ba}$ denotes the linear projection onto the
column space of the observed anchor variable matrix $\ba$. If the anchor
variables $A$ have a linear effect onto $X$, $Y$ and the hidden confounders
$H$, it is reasonable to consider the estimator with the linear
projection operator $\Pi_{\ba}$. We will give some justification for it in
Section \ref{subsec.heuristics}.  

It is straightforward to see that the objective function can be represented as 
\begin{eqnarray*}
G(f) = \|W(Y - f)\|_2^2/(2n),\ \ W = W_{\gamma} = I - (1 - \sqrt{\gamma}) \Pi_{\ba}.
\end{eqnarray*}

\subsection{Anchor Boosting: a ``regularized'' approximation of the estimator}
The question is how to compute or approximate the estimator in
\eqref{estimator}. We aim here for a solution where standard existing
software can be used, 

Our proposal is to use boosting. For this, we consider
the negative gradient 
\begin{eqnarray*}
- \frac{\partial}{\partial f} G(f) = W^2(Y-f)/n
\end{eqnarray*}
and pursue iterative fitting of the negative gradient. The negative
gradient fitting is done with a pre-specified base learner (or ``weak
learner''): it is a regression estimator $\hat{f}_{U,X}$ based on
input data $(U,X)$ with $U$ denoting a response vector (e.g. $U = Y$
corresponds to the estimator applied to the original data). This is the standard
recipe of gradient boosting \citep{brei99,fried01,pbyu03,BuhlmannHothorn06}
and the method is summarized in Algorithm \ref{alg.AB-boosting}. 
\begin{algorithm}[!htb]
\begin{algorithmic}[1]
\STATE Initialize with $f^{[0]} \equiv 0$. Set $m = 0$.
\STATE Increase $m$ by 1: $m \leftarrow m+1$.
\STATE Compute the pseudo-response $\tilde{Y} = W^2(Y - f^{[m-1]})/n$ which
  equals the negative gradient vector evaluated at $f^{[m-1]}$. 
\STATE Compute the regression function estimator $\hat{f}_{\tilde{Y},X}$
from the base learner and up-date
\begin{eqnarray*}
f^{[m]} = f^{[m-1]} + \nu \cdot \hat{f}_{\tilde{Y},X},
\end{eqnarray*}
where $0 < \nu < 1$ is a pre-specified parameter. The default value is $\nu
= 0.1$. 
\STATE Repeat steps 2-4 until reaching a stopping iteration
$m_{\mathrm{stop}}$. 
\end{algorithmic}
\caption{Anchor Boosting algorithm.}\label{alg.AB-boosting}
\end{algorithm}
The choice of the stopping iteration $m_{\mathrm{stop}}$ is discussed in
Section \ref{subsec.stopping}. The stopping iteration is a regularization
parameter: it is governing a bias-variance trade-off, on top of the
regularization of causal regularization from anchor regression which is
encoded in the matrix $W = W_{\gamma}$. 

From another view point for regularization, we can think of the regression
estimator $\hat{f}_{\tilde{Y},X}$ as an 
operator $B$ (when evaluated at the observed $X$; it is a ``hat'' operator):
\begin{eqnarray*}
B: (\tilde{Y},X) \mapsto \hat{f}_{\tilde{Y},X}(\cdot).
\end{eqnarray*}
Then, it is straightforward to see that 
\begin{eqnarray*}
f^{[m]}(X_1),\ldots ,f^{[m]}(X_n) = \left(I - (I - W^2B)^m\right)Y,
\end{eqnarray*}
that is, the boosting operator at iteration $m$ is equal to $\left(I - (I -
  W^2B)^m\right)$. If $W^2B$ has a suitable norm being strictly $<1$, then
there is geometrical convergence to the identity which would fit the data
$Y$ perfectly. This indicates, that we should stop the boosting procedure
to avoid overfitting. However, especially for large values of $\gamma$ in $W
= W_{\gamma}$, the norm of $W^2B$ will be larger than one and geometrical
contraction, i.e. overfitting, to the response vector $Y$ will not happen. 

\subsubsection{Some criteria for choosing the stopping
  iteration.}\label{subsec.stopping}

In connection with a Random Forests \citep{brei01} learner for the estimator
$\hat{f}_{\tilde{Y},X}$ in step 5 of Algorithm \ref{alg.AB-boosting} we
found that we can choose the stopping 
iteration $m$ such as to minimize the objective function $\|W_{\gamma}(Y -
f^{[m]})\|_2^2$ or even overshooting the minimum by say 10\%. Formally, the
two stopping rules are:
\begin{eqnarray}
& &m_{\mathrm{stop}} = \argmin_m \|W_{\gamma}(Y - f^{[m]})\|_2^2,\label{stop1}\\
& &m_{\mathrm{stop}} = \mbox{argmax}_m \|W_{\gamma}(Y - f^{[m]})\|_2^2\nonumber\\
& &\hspace*{8.5mm} \mbox{such that}\
    \|W_{\gamma}(Y - f^{[m]})\|_2^2 \le 1.1 \min_m  \|W_{\gamma}(Y -
    f^{[m]})\|_2^2.\label{stop2} 
\end{eqnarray}

In general, we propose a rule based on the following observation. 
Consider the population version of $W = W_{\gamma}$ and denote it by 
\begin{eqnarray*}
R_{\gamma} = \mathrm{Id} - (1 - \sqrt{\gamma}) P_A,
\end{eqnarray*}
where $P_A(\cdot)$ denotes the best linear projection onto $A$. That is,
\begin{eqnarray*}
P_A(Z) = (\alpha^0)^T A,\ \alpha^0 = \mbox{argmin}_{\alpha} \EE[(Z -
  \alpha^T A)^2] = \mathrm{Cov}(A)^{-1} \mathrm{Cov}(A,Z).
\end{eqnarray*}
Thus, $R_{\gamma}$ is a function of $A$ and random.  

The population version of the optimization is 
\begin{eqnarray*}
\argmin_{f \in C_X} \EE[(R_{\gamma}(Y - f(X)))^2] = \argmin_{f \in C_X} \EE[(R_{\gamma}Y - R_{\gamma}f(X))^2], 
\end{eqnarray*} 
due to linearity of $R_{\gamma}$; here $C_X$ denotes the class of
measurable functions of $X$. We decompose this problem into two parts:
\begin{eqnarray*}
& &\EE[(R_{\gamma}Y - R_{\gamma}f(X))^2] = \EE[(R_{\gamma} Y -
  g_{\mathrm{opt}}(X,A))^2] + \EE[(g_{\mathrm{opt}}(X,A) - R_{\gamma}f(X))^2],\\
& &g_{\mathrm{opt}} = \EE[R_{\gamma}Y|X,A].
\end{eqnarray*}
This motivates a finite sample criterion guarding against
overfitting. Whenever we estimate 
  $f(\cdot)$ by $\hat{f}(\cdot)$ with boosting as in Algorithm
  \ref{alg.AB-boosting}, the residual sum of squares $\|W_{\gamma} (Y -
  \hat{f}(X))\|_2^2$ should be at least as 
  large as the residual sum of squares of $\|W_{\gamma}Y -
  \hat{g}_{\mathrm{opt}}(X,A)\|_2^2$ of any good and reasonably tuned
  machine learning estimator $\hat{g}_{\mathrm{opt}}$. That is, we should
  choose the number 
  of boosting iterations such that it minimizes the objective function
  under the given constraint: 
\begin{eqnarray}\label{stop3}
& &m_{\mathrm{stop}} = \min_m \|W_{\gamma}(Y - f^{[m]}(X))\|_2^2\nonumber\\
& &\hspace*{4mm}  \mbox{such
  that}\ \|W_{\gamma}(Y - f^{[m]}(X))\|_2^2 \ge \|W_{\gamma}Y -
  \hat{g}_{\mathrm{opt}}(X,A)\|_2^2.
\end{eqnarray}
This guards against overfitting and avoids choosing a too large boosting
iteration. We could also modify the rule to choose $m$ as in \eqref{stop2}
under the constraint that $\|W_{\gamma}(Y - f^{[m]}(X))\|_2^2 \ge \|W_{\gamma}Y -
  \hat{g}_{\mathrm{opt}}(X,A)\|_2^2$. 

\subsubsection{Random Forests learner with a linear model
  component.}\label{subsec.LMRFlearner} 

Besides using Random Forests as a base learner $\hat{f}_{U,X}$ in the
Anchor-Boosting Algorithm \ref{alg.AB-boosting}, we consider also a modification
which fits a linear model first and applies 
Random Forests on the resulting residuals. This modification is denoted by
``LM+RF'', standing for Linear Models+Random Forests. The LM+RF procedure
is built on the 
idea that the linear model part is the ``primary part'' 
and the remaining nonlinearities are then estimated by Random Forests.
This base leaner is able to cope well with estimating partial linear functions. 

The ``LM+RF'' algorithms is defined as follows. Given a response variable $Y$
and some covariates $X$:
\begin{enumerate}
\item Fit a linear model of $Y$ versus $X$, by default including an
  intercept. The fitted (linear) regression function is denoted by
  $\hat{f}_1$. 
\item Compute the residuals from step 1, denoted by $R$. Fit a Random
  Forests of $R$ (being now the response variable) versus $X$: the fitted
  regression function is denoted by $\hat{f}_2$. 
\item The final estimator is $\hat{f}_1 + \hat{f}_2$. 
\end{enumerate}
The LM+RF base learner is typically outperforming plain Random Forests
if the underlying regression function is an additive combination of a
linear and a nonlinear function.

\subsubsection{Plug-in of any ``machine learning'' algorithm.} 

Obviously, any ``machine learning'' regression technique can be used as
base learner in the Anchor Boosting Algorithm \ref{alg.AB-boosting}. In
addition, also the stopping rule in \eqref{stop3} involving an estimator
$\hat{g}_{\mathrm{opt}}(X,A)$ can be used with any reasonable regression
algorithm. 

\subsection{Some empirical results}\label{subsec.empirres}

We consider the following structural equation model for the in-sample data
used for training with sample size $n = 300$.  
\begin{eqnarray}\label{SEM1}
& &A \sim {\cal N}_2(0,\mathrm{I}),\nonumber\\
& &H \sim {\cal N}_1(0,1),\nonumber\\
& &X_j = A_1 + A_2 + 2H + \eps_{X,j}\ (j = 1\ldots ,p),\ \eps_{X} \sim
    {\cal N}_{10}(0,0.5^2 \cdot \mathrm{I}),\nonumber\\
& &Y = f(X_2,X_3) - 2 A_1 + 3H  + \eps_Y,\ \eps_Y \sim {\cal
    N}_1(0,0.25^2),
\end{eqnarray}
where $A,H,\eps_X,\eps_Y$ are jointly independent. The dimensions
of the variables are $\dim(A) = 2$, $\dim(H) = 1$, $\dim(X) = 10$ and
$\dim(Y) = 1$. \begin{figure}[!b]
\begin{center}
\includegraphics[scale=0.7]{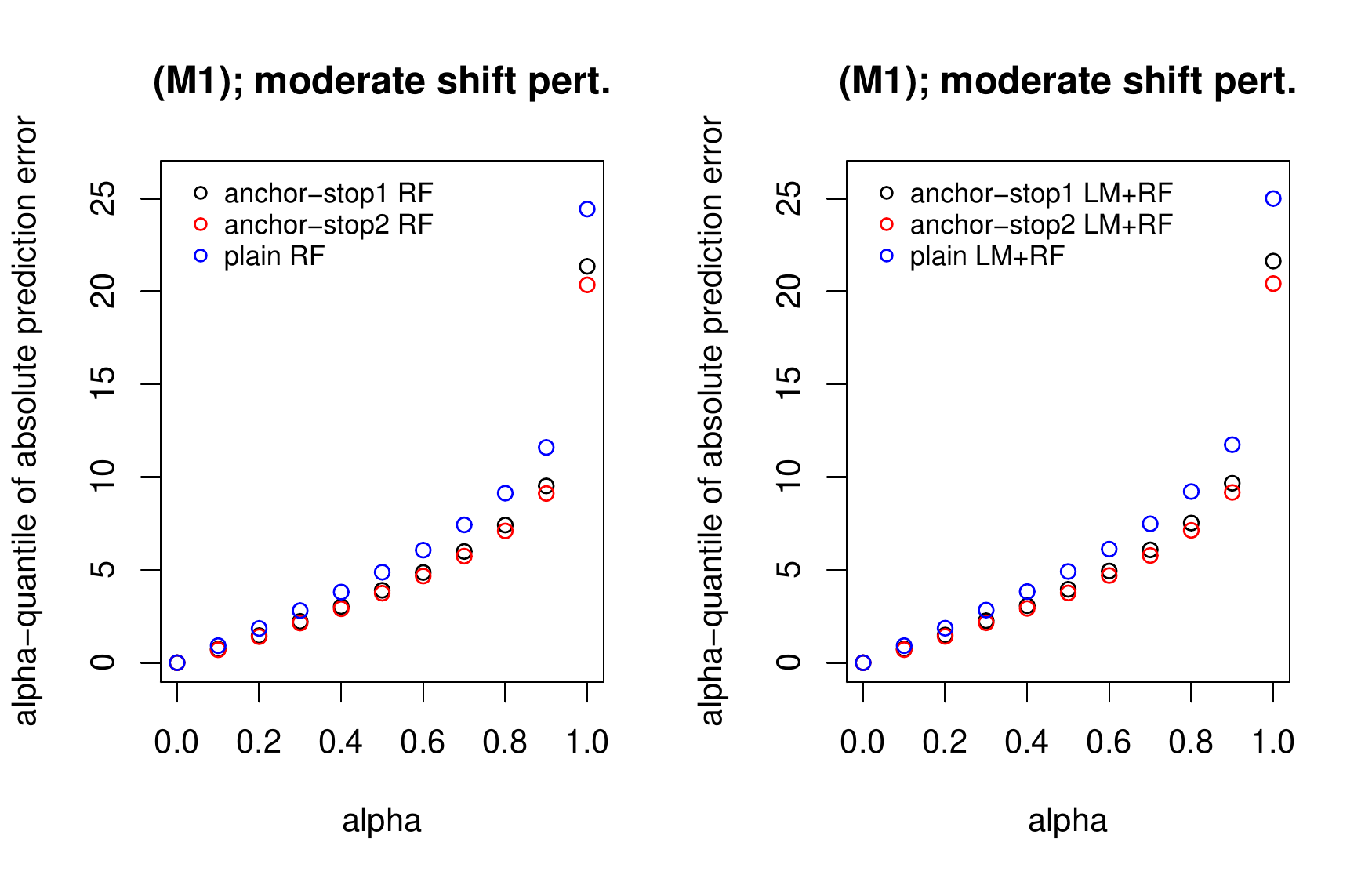}
\includegraphics[scale=0.7]{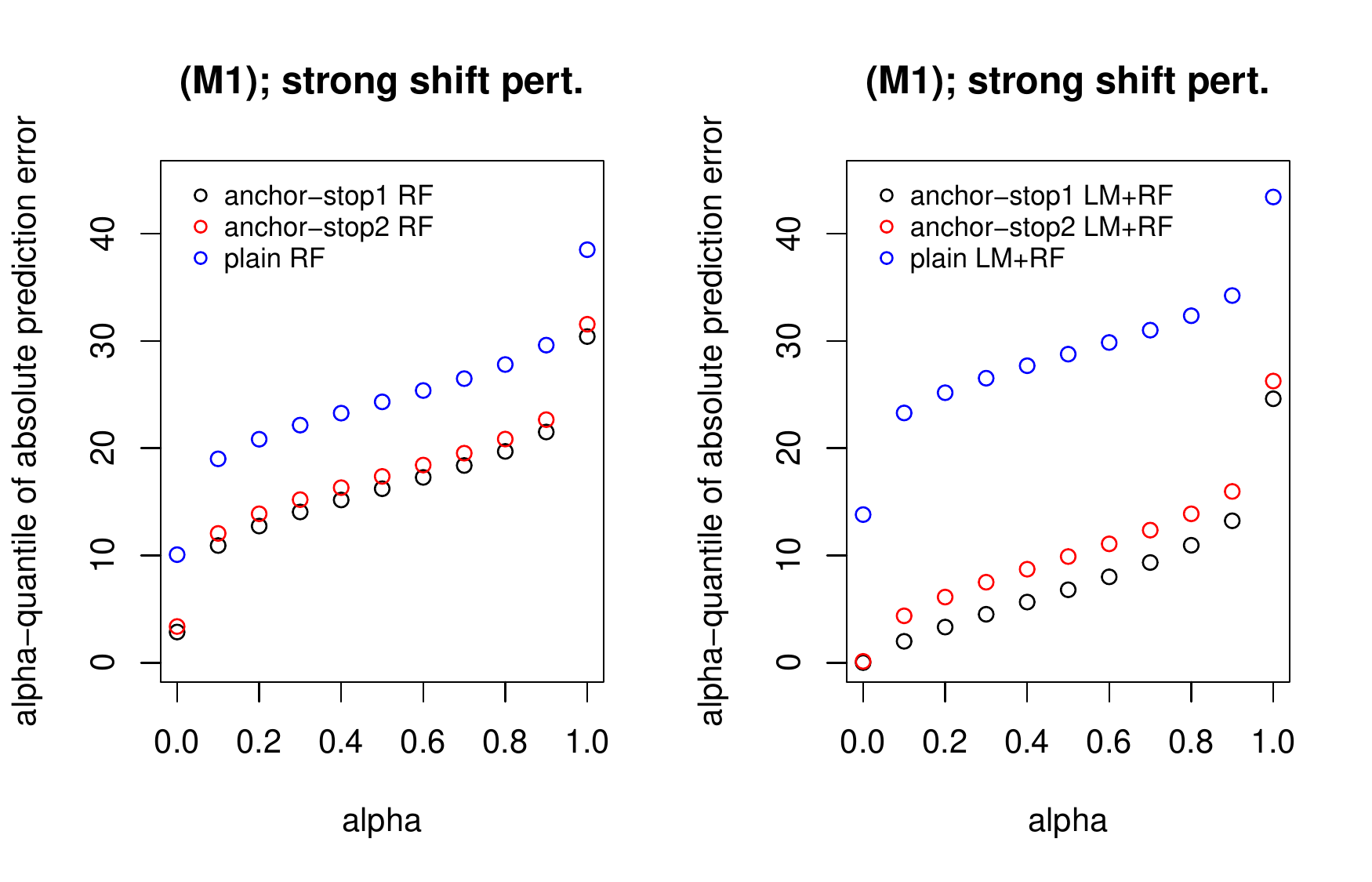}
\caption{Empirical $\alpha$-quantiles of $|Y_{\mathrm{out},i} -
  \hat{Y}_{\mathrm{out},i}|$ for $i = 1,\ldots, n_{\mathrm{out}} = 2000$,
  averaged over 100 independent simulation runs. Model (M1) and moderate
  shift perturbations (i) for $A_{\mathrm{out}}$ (top) and strong shift
  perturbations (ii) for $A_{\mathrm{out}}$ (bottom). The Anchor Boosting
  algorithm is always used with $\gamma = 7$, with Random
Forests (left) and with Linear Model + Random Forests
(right), and with the two stopping criteria from \eqref{stop1} (stop1) and \eqref{stop2} (stop2).}\label{fig.nonlin1}  
\end{center}
\end{figure}
For the function $f(\cdot)$, we consider the following two
models:
\begin{description}
\item[(M1)] $I(X_2 \le 0) + I(X_2 \le -0.5) I(X_3 \le 1)$,
\item[(M2)] $X_2 + X_3 + I(X_2 \le 0) + I(X_2 \le -0.5) I(X_3 \le 1)$.
\end{description}
The model (M1) has no linear term while (M2) does have one; for both
models, the number of active variables is $2$. 
%
%
%
The out-of-sample data is generated according to the same structural
equation model as in \eqref{SEM1} but with two different perturbations for
the anchor variables, denoted by $A_{\mathrm{out}}$. We consider the following:
\begin{description}
\item[(i)] moderate shift perturbation:
\begin{eqnarray*}
& &\mu \sim {\cal N}_{\mathrm{nout}}(1,2^2\mathrm{I}),\\
& &A_{\mathrm{out}} \sim {\cal N}_{\mathrm{nout}}(\mu,\mathrm{I}),
\end{eqnarray*}
where $\mathrm{nout} = 2000$ denotes the number of out-of-sample observations. 

\item[(ii)] strong shift perturbation: 
\begin{eqnarray*}
& &\mu \sim {\cal N}_{\mathrm{nout}}(10,\mathrm{I}),\\
& &A_{\mathrm{out}} \sim {\cal N}_{\mathrm{nout}}(\mu,\mathrm{I}),
\end{eqnarray*}
where $\mathrm{nout} = 2000$ denotes the number of out-of-sample observations. 
\end{description}

\begin{figure}[!htb]
\begin{center}
\includegraphics[scale=0.7]{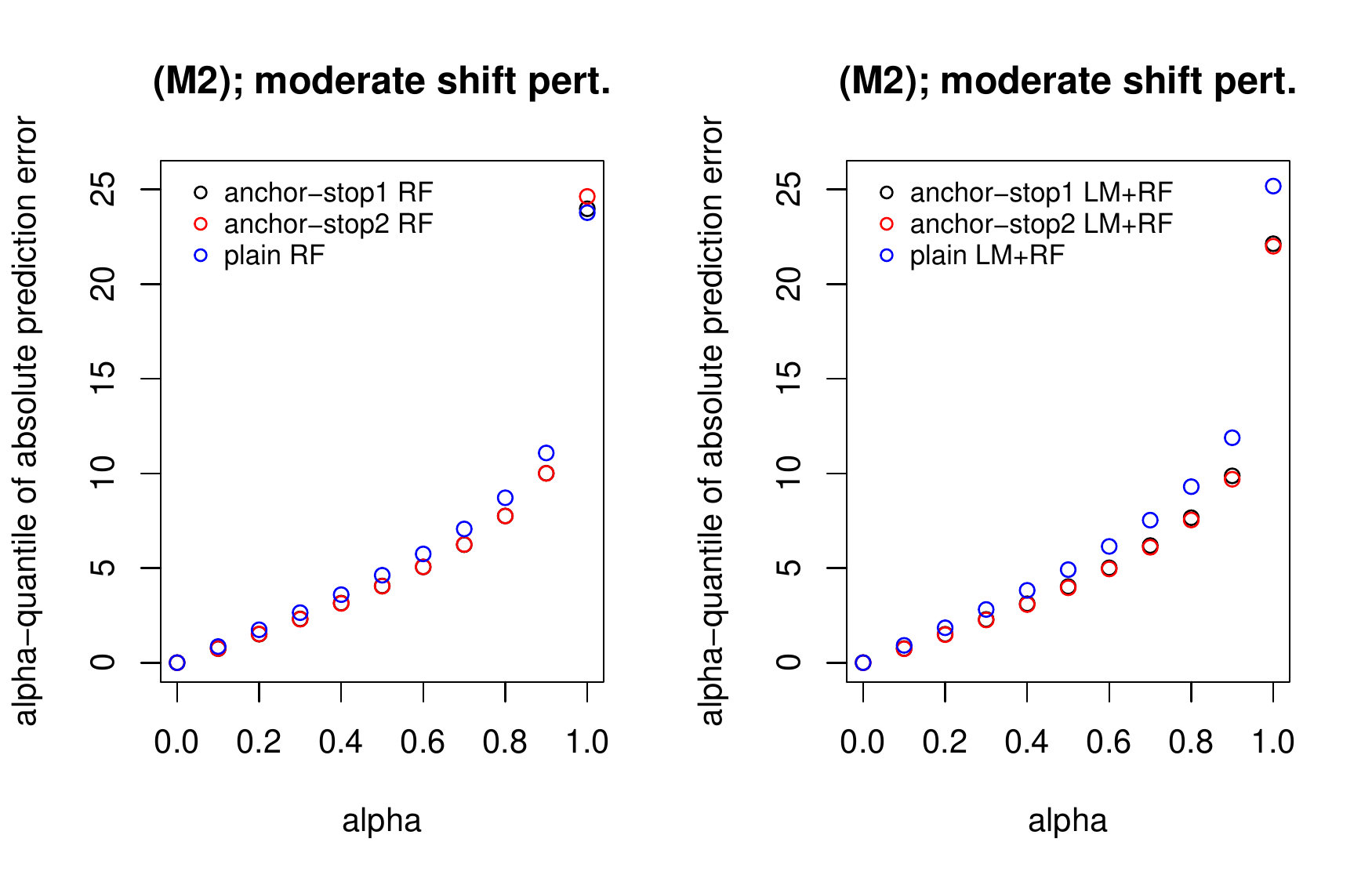}
\includegraphics[scale=0.7]{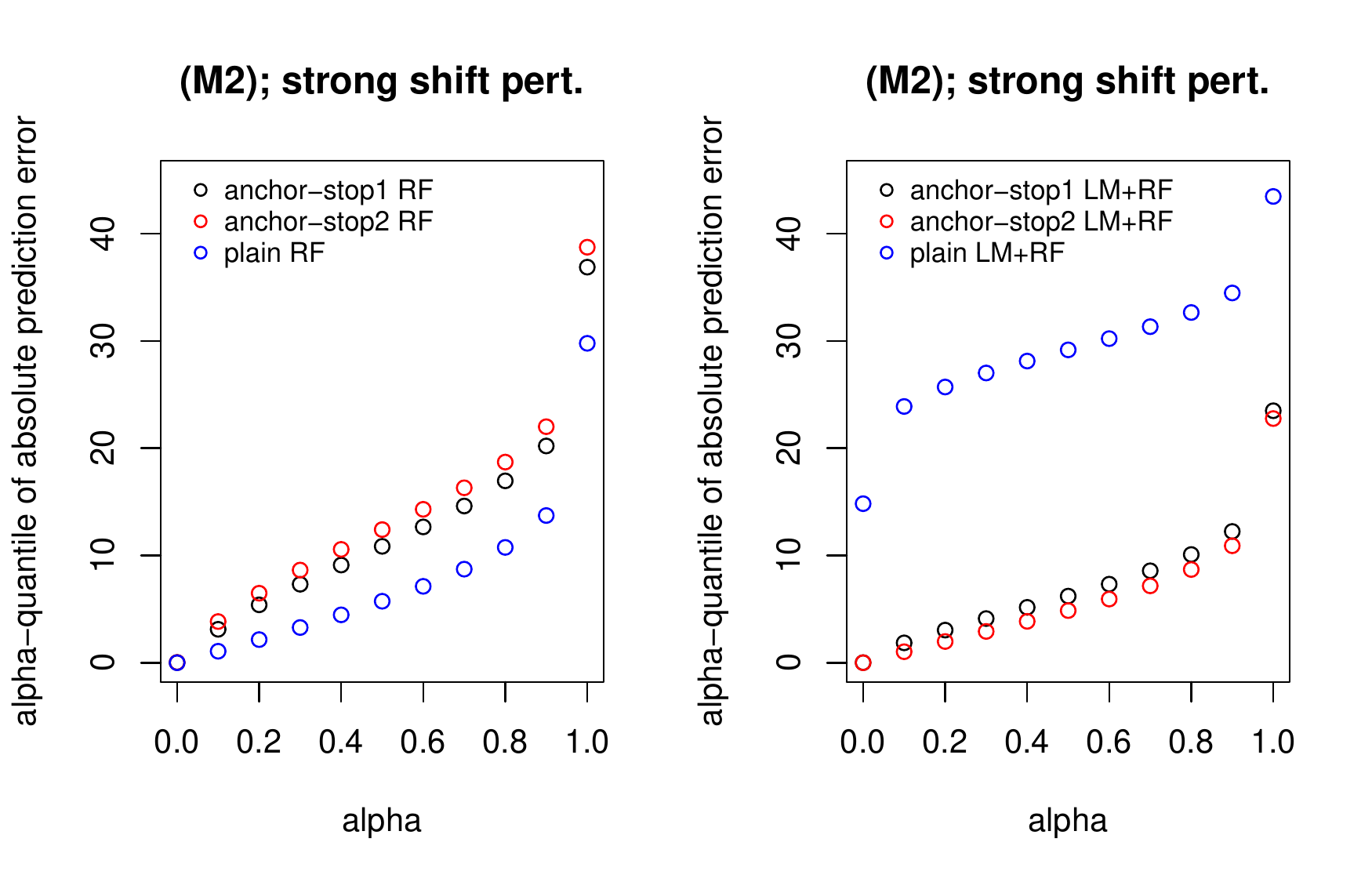}
\caption{Empirical $\alpha$-quantiles of $|Y_{\mathrm{out},i} -
  \hat{Y}_{\mathrm{out},i}|$ for $i = 1,\ldots, n_{\mathrm{out}} = 2000$,
  averaged over 100 independent simulation runs. Model (M2) and moderate
  shift perturbations (i) for $A_{\mathrm{out}}$ (top) and strong shift
  perturbations (ii) for $A_{\mathrm{out}}$ (bottom). The Anchor Boosting
  algorithm is always used with $\gamma = 7$, with Random
Forests (left) and with Linear Model + Random Forests
(right), and with the two stopping criteria from \eqref{stop1} (stop1) and \eqref{stop2} (stop2).}\label{fig.nonlin2}
\end{center}
\end{figure}

We report some performances of Anchor Boosting with Random Forests, Anchor
Boosting with the LM+RF learner from Section \ref{subsec.LMRFlearner} and
plain Random Forests in Figures \ref{fig.nonlin1}-\ref{fig.nonlin2}. We do
not tune the 
parameter $\gamma$ and consider only the choice $\gamma = 7$. The
performance measures are empirical $\alpha$-quantiles of the out-of-sample
predictions $|Y_{\mathrm{out},i} - \hat{Y}_{\mathrm{out},i}|$ for a range
of different $\alpha$-values. 
In terms of quantitative numbers, the relative performance gain of Anchor
Boosting with stopping as in \eqref{stop2} (``stop2'') over the
corresponding plain Random Forests algorithm is given in Table \ref{tab.1}.
\begin{table}[!htb]
\begin{center}
\begin{tabular}{l|c}
model \& learner & performance gains at $\alpha \in \{0.5,0.8,1\}$\\
\hline
(M1); moderate shift \& RF & 23.2\%, 22.3\%, 16.7\%\\
(M1); moderate shift \& LM+RF & 23.5\%, 22.7\%, 18.3\%\\
(M1); strong shift \& RF & 28.6\%, 25.0\%, 18.0\%\\
(M1); strong shift \& LM+RF & 65.6\%, 57.1\%, 39.5\%\\
(M2); moderate shift \& RF & 12.5\%, 11.1\%,\ -3.7\%\\
(M2); moderate shift \& LM+RF & 19.5\%, 18.9\%, 12.6\%\\
(M2); strong shift \& RF &  -100.8\%, -73.9\%, -30.1\%\\
(M2); strong shift \& LM+RF & 83.3\%, 73.4\%, 47.6\%
\end{tabular}
\end{center}
\caption{Relative performance gain of empirical $\alpha$-quantiles of absolute
  out-of-sample prediction errors (see also captions in Figures
  \ref{fig.nonlin1}-\ref{fig.nonlin2}) of Anchor Boosting with $\gamma = 7$
  and stopping from \eqref{stop2} over the corresponding plain base learner. The base learners are Random Forests (RF) and Linear Model + Random Forests (LM+RF).}\label{tab.1}
\end{table}

The performance gain with Anchor Boosting is substantial, with the only
exception of the case (M2) with strong shift perturbations (ii) and the Random
Forest base learner (RF). This is a
situation where a RF learner is ``mis-specified'' since it cannot capture
well the linear function part in model (M2): it is particularly harmful with
strong shift perturbations where a strong shift in the anchor variables $A_{\mathrm{out}}$
results in strong shifts of the covariates and through the linear function
also in $Y$. The RF learner cannot capture such strong shifts
well; when using the much better Linear Model + Random Forests (LM+RF)
learner, the performance gain of Anchor Boosting is massive.
When comparing Anchor Boosting with LM+RF to plain Random Forests (RF), the
relative gain over plain RF at $\alpha \in \{0.5,0.8,1\}$ in the model 
(M2) with strong strong shift perturbations is  
\begin{eqnarray*}
15.2\%, 19.2\%, 23.6\%
\end{eqnarray*}
which is again in clear favor of Anchor Boosting with the LM+RF learner.

Even if the model has no linear function (as in model (M1)), it seems to
pay-off to use the LM+RF learner in presence of strong shift
interventions. A possible reason is that extrapolation for large $X$-values
is not easily possible with Random Forests.

\subsection{Variable importance}\label{subsec.varimp}
Of particular interest is the notion of variable importance: in connection
with (nonlinear) anchor regression, it is a more causally oriented measure
than a plain variable importance measure in standard (nonlinear)
regression. See also the term of ``diluted causality'' in Section \ref{subsec.diluted-causality}.

We suggest a variable importance measure based on permutation, following
Breiman's proposal for Random Forests \citep{brei01}. Unlike Breiman's
original proposal which involves the out-of-bag observations occurring in
Random Forests, we work with the training data only. It seems to work as
long as the estimated anchor regression function is reasonably regularized
and does not overfit. 

Consider an estimated anchor regression function
$\hat{f}_{\mathrm{anchor}}(\cdot): {\cal X} \to {\cal Y}$. For $j \in
\{1,\ldots ,p\}$, permute the $j$th covariate $X_j$ (permute the sample
indices) and denote this permuted variable by
$X_{\mathrm{perm}(j)}$. We define the variable, for $j \in \{1,\ldots,p\}$
\begin{eqnarray*}
\tilde{X}_{\mathrm{perm}(j)}^{(i)} = (X_1^{(i)},\ldots
  ,X_{j-1}^{(i)},X_{\mathrm{perm}(j)}^{(i)},X_{j+1}^{(i)},\ldots ,X_p^{(i)})^T,\
  i=1,\ldots ,n,
\end{eqnarray*}
where the $j$th component is permuted relative to the observed variable
$X^{(i)}$. 
We then compute the residual sum of squares
\begin{eqnarray*}
\mathrm{RSS}_j = n^{-1} \sum_{i=1}^n (Y^{(i)} -
  \hat{f}_{\mathrm{anchor}}(\tilde{X}_{\mathrm{perm}(j)}^{(i)}))^2.
\end{eqnarray*}
The importance measure is the relative increase in of $\mathrm{RSS}_j$ in
comparison to the standard $\mathrm{RSS} = n^{-1} \sum_{i=1}^n (Y^{(i)} -
  \hat{f}_{\mathrm{anchor}}(X^{(i)}))^2$:
\begin{eqnarray}\label{imp1}
\mathrm{Imp}_j = \frac{\mathrm{RSS}_j - \mathrm{RSS}}{\mathrm{RSS}}.
\end{eqnarray}
As an alternative, we also consider the median absolute loss instead of the
residual sum of squares:
\begin{eqnarray}\label{imp2}
& &\mathrm{Imp}_{\mathrm{med},j} = \frac{\mathrm{h}_j -
  \mathrm{h}}{\mathrm{h}},\nonumber\\
& &h_j = \mbox{sample median} \{|Y^{(i)} -
    \hat{f}(\tilde{X}_{\mathrm{perm}(j)}^{(i)})|;\ i=1,\dots ,n\},\nonumber\\
& &h =
    \mbox{sample median} \{|Y^{(i)} - \hat{f}(X^{(i)})|;\ i=1,\dots ,n\}.
\end{eqnarray}

For empirical results we note that the models (M1) and (M2) from Section \ref{subsec.empirres} are very difficult in terms of
identifying important variables. In fact, with the latter models, the
correlation among the covariates $X$ is extremely strong: the average
absolute value of the off-diagonal elements of the empirical (in-sample)
correlation matrix is found to be 
\begin{eqnarray*}
\frac{1}{p(p-1)} \sum_{j \neq k} |\widehat{\mbox{Corr}}(X_j,X_k)| = 0.97
\end{eqnarray*}
for a representative sample.
\begin{figure}[!t]
\begin{center}
\includegraphics[scale=0.5]{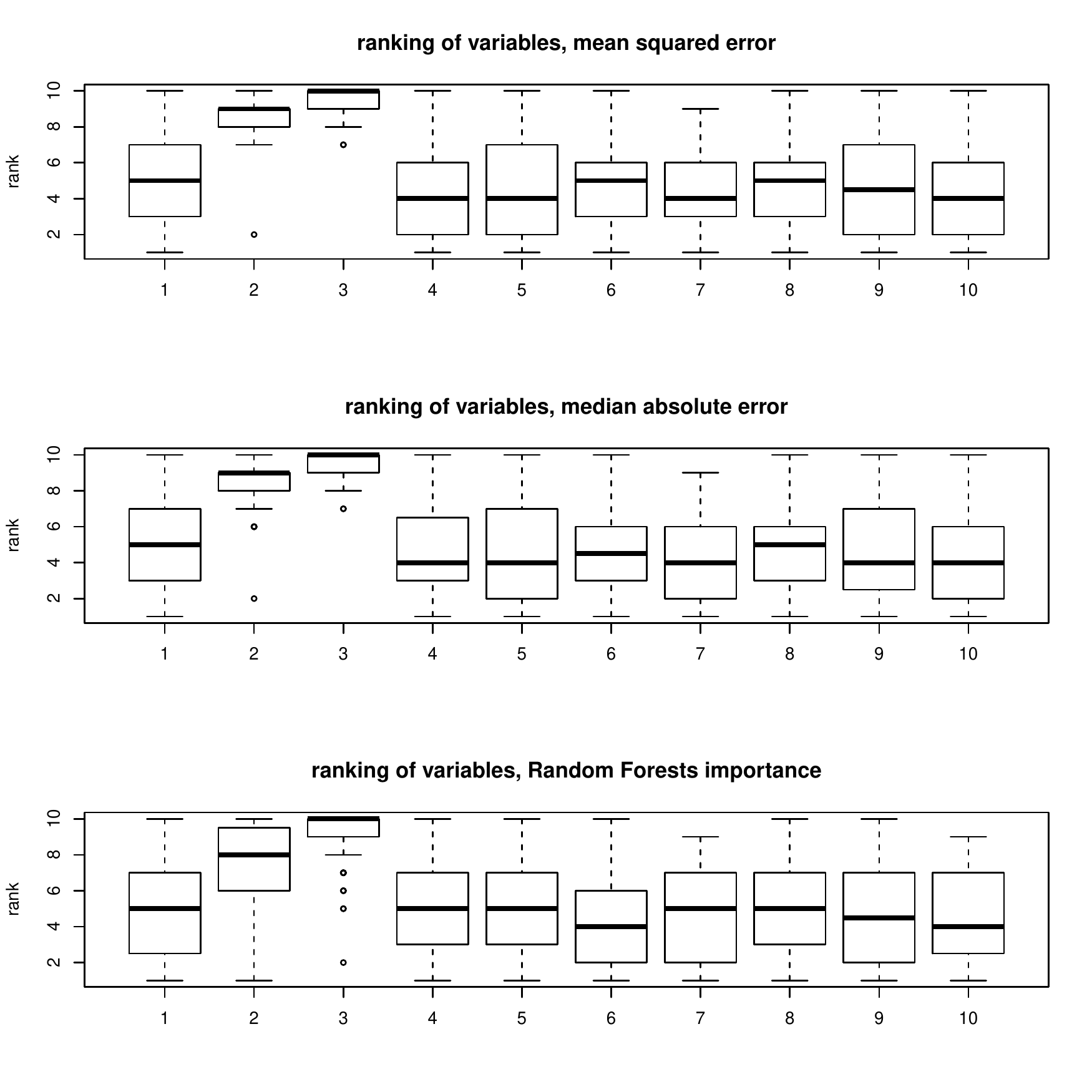}
\caption{Boxplots of the ranks of variable importance (rank 1 lowest
  priority for variable importance, rank 10 highest priority), based on 100
  independent simulations of model (M3). The different boxes correspond to
  the different variable indices $j \in \{1,\ldots ,10\}$. Anchor Boosting
  with LM+RF and $\gamma = 7$: measure \eqref{imp1} (top), measure
  \eqref{imp2} (middle); 
  variable importance of standard Random Forests measuring the increase in
  residual sum of squares based on out-of-bag observations
  (bottom).}\label{fig.varimportance}  
\end{center}
\end{figure}
We modify to the following model: 
\begin{description} 
\item[(M3)] The structural equation model is: 
\begin{eqnarray*}
& &A \sim {\cal N}_2(0,\mathrm{I}),\\
& &H \sim {\cal N}_1(0,1),\\
  & &\Gamma\ \mbox{a $2 \times 10$ matrix with i.i.d. ${\cal N}(0,1)$ entries},\\
& &X_j = A^T \Gamma_{\bullet j} + H + \eps_{X,j}\ (j = 1\ldots ,p),\ \eps_{X} \sim
    {\cal N}_{10}(0,\mathrm{I}),\\
& &Y = f(X_2,X_3) - 2 A_1 + 3H  + \eps_Y,\ \eps_Y \sim {\cal
    N}_1(0,0.25^2),
\end{eqnarray*}
where $A,H,\eps_X,\eps_Y,\gamma$ are jointly independent and 
\begin{eqnarray*}
f(x_2,x_3) = x_2 + x_3 + I(x_2 \le 0) + I(x_2 \le -0.5) I(x_3 \le 1).
\end{eqnarray*}
The dimensions
of the variables are $\dim(A) = 2$, $\dim(H) = 1$, $\dim(X) = 10$ and
$\dim(Y) = 1$. 
\end{description}
We consider again sample size $n = 300$ and out-of-sample size
$n_{\mathrm{out}} = 2000$ (for variable importance, we do not need this). 
In terms of the empirical performance for prediction, the analogue of Table
\ref{tab.1} is as follows. 
\begin{center}
\begin{tabular}{l|c}
model \& learner & performance gains at $\alpha \in \{0.5,0.8,1\}$\\
\hline
(M3); strong shift \& RF & \ 9.6\%,\ \ \ 8.4\%, 6.3\%\\
(M3); strong shift \& LM+RF & 28.5\%, 18.5\%, 1.5\%\\
\end{tabular}
\end{center}

For variable importance, the measures $\mathrm{Imp}_j$ and
$\mathrm{Imp}_{\mathrm{med},j}$ are displayed in Figure \ref{fig.varimportance}.
Since the variables $X_2$ and $X_3$ are the only active variables in the
true function $f(X)$ in model (M3) we conclude that Anchor Boosting with
LM+RF does a substantially better job for quantifying variable importance
than standard Random Forests. Note that, as in Section
\ref{subsec.empirres}, there was no tuning for the parameter $\gamma$ which
was set to the value $\gamma = 7$.  

\subsection{Some arguments why a simple linear projection $\Pi_{\ba}$ is sufficient}\label{subsec.heuristics}

We present here some arguments under what conditions a linear projection
$\Pi_{\ba}$ in the estimator in \eqref{estimator} is reasonable even in
presence of a nonlinear function $f(\cdot)$.  

We consider the situation when the regularization parameter
$\gamma \to \infty$. Then, in the population version, the regularization
enforces a solution $f(\cdot)$ with $\mbox{Corr}(A, Y - f(X)) = 0$. We thus
consider the set 
\begin{eqnarray*}
I = \{f(\cdot);\ \EE[A(Y - f(X))] = 0\ \mbox{and}\ \EE[Y - f(X)] = 0\}.
\end{eqnarray*}
We then have the following result.
\begin{prop}\label{prop1}
Consider a nonlinear anchor regression model as in
\eqref{nonlin.anchor.model}.
We assume that $\Cov(A)$ is positive definite. Assume that $\EE[Y - f(X)|A] =
\mu_f + \alpha_f^T A$ is a linear function of $A$ with $\mu_f \in \R$ and
$\alpha_f$ a $r \times 1$ vector. 
Then, for any $f \in I$ we have for every $\mathrm{do}$-perturbation on $A$
with $\mathrm{do}(A = a)$ \citep[cf.]{pearl00}, where the value $a$ is
deterministic or random:  
\begin{eqnarray*}
\EE[Y^a - f(X^a)] = \mu_f =\ \mbox{constant w.r.t.}\ a.
\end{eqnarray*}
Here, $Y^a$ and $X^a$ denote the random variables corresponding to the
do-perturbation $\mathrm{do}(A = a)$. 
\end{prop}
A proof is given at the end of the section. 
Proposition \ref{prop1} leads to invariance of the first moment of the
residuals $Y^a - f(X^a)$ but is not saying anything on higher moments or
the invariance of the distribution as in Proposition \ref{prop.shiftinv}. 

For nonlinear $f(\cdot)$, the conditional expectation $\EE[Y - f(X)|A]$ is
typically non-linear in $A$, thus violating the assumption of Proposition
\ref{prop1}. There is one very notable an relevant exception though: for
discrete anchor variables $A \in {\cal A}$ and discrete space ${\cal A} =
\{1,\ldots ,m\}$ with labels $1,\ldots ,m$, we can always write 
\begin{eqnarray*}
\EE[Y - f(X)|A = a] = \mu_f + \alpha_f^T A
\end{eqnarray*}
where $A$ is now with dummy-encoding with  $r = (m-1)$-dimensional
representation
Positive definiteness of $\Cov(A)$ is
ensured by assuming $0 < \PP[A = k] < 1$ for all $k$. This leads to the
following consequence:  
\begin{corr}\label{corr1}
  Assume discrete anchor variables $A$ with dummy encoding.
  We further assume that $0 < \PP[A = k] < 1$ for all $k$. Then, for any $f
  \in I$ we have for every  $\mathrm{do}$-perturbation on $A$  
with $\mathrm{do}(A = a)$, where the value $a$ is deterministic or random: 
\begin{eqnarray*}
\EE[Y^a - f(X^a)] = \mu_f =\ \mbox{constant w.r.t.}\ a.
\end{eqnarray*}
\end{corr}
We note that a $\mathrm{do}$-perturbation on $A$ would simply change the
value of the dummy-encoding: say $\mathrm{do}(A_j = 5)$ would imply a
10-fold effect if the observed $A_j = 0.5$. 

We point out that we do not need a specific additive form in the
structural equation model. However, Corollary \ref{corr1} is only
interesting if the set $I$ is non-empty. This is ensured if the structural
equation for $Y$ is additive in $X$ and $H$ and the anchor $A$ is an
instrument only influencing $X$, i.e., for a broad class of nonlinear
instrumental variable regression models which satisfies:
\begin{eqnarray*}
& &Y \leftarrow f(X) + g(H,\eps_Y),\\
& &X \leftarrow h(A,H,\eps_X),
\end{eqnarray*}
where $A, H, \eps_X, \eps_Y$ are mutually independent exogenous variables
(source nodes in the graph), 
Obviously, $Y - f(X)$ is then independent of $A$ and therefore $f \in
I$. We do not elaborate more under what conditions $I$ is non-empty: in the
case of linear structural equations we note the result in Proposition \ref{prop.shiftinv}. 

For non-discrete anchor variables and where the conditional expectation
$\EE[Y - f(X)|A] = \mu_f + \alpha_f^T A + A^T \beta_f A + \ldots $, with
$\beta_f$ a $r \times r$ 
matrix, is nonlinear in $A$, the nonlinear anchor boosting algorithm leads
to invariance of the linearized part:
\begin{eqnarray*}
\EE[Y^a - f(X^a)] \approx \mu_f + \EE[a^T \beta_f a] + \ldots
\end{eqnarray*}
where the expression does not involve dependence on $A$ through the linear
part $\alpha_f^T a$. A 
linear approximation is more likely to be good if $A$ has only a linear
influence on $X$, $H$ and $Y$ and if $f(\cdot)$ is not too far away from a
linear function. Thus, we would conclude that the nonlinear anchor boosting
estimator leads to good predictive robustness if the direct effects of $A$
are linear and the nonlinearity enters via a nonlinear function
$f(\cdot)$ in the structural equation for $Y$. If the conditional
expectation $\EE[Y - f(X)|A]$ becomes highly non-linear, one would need a
different penalization which could be of the form such as 
\begin{eqnarray*}
\max_{g \in {\cal G}} \EE[g(A)(Y - f(X))],
\end{eqnarray*}
where ${\cal G}$ is a suitable class of functions.

\medskip
\paragraph{Proof of Proposition \ref{prop1}.}
Since $A$ is source node, the do-perturbation on $A$ is the same as
the conditional expectation:
\begin{eqnarray*}
\EE[Y - f(X)|A = a] = \EE[Y - f(X)|\mathrm{do}(A=a)] = \EE[Y^a - f(X^a)],
\end{eqnarray*}
where the last equality is just a definition. 

We have for $f \in I$ that 
\begin{eqnarray*}
\EE[A(Y - f(X))] = 0.
\end{eqnarray*}
Therefore by linearity of the conditional expectation we obtain
\begin{eqnarray*}
0 &=& \|\EE[A (Y - f(X))]\|_2^2  = \|\EE[A \EE[Y - f(X)|A]]|_2^2 = \| \EE[A (\mu_f + \alpha_f^T A)\|_2^2 \\
& = &\|\EE[A \alpha_f^T (A -
      \EE[A])]\|_2^2 = \|\EE[(A - \EE[A]) \alpha_f^T (A -
      \EE[A])]\|_2^2\\
& =&\|\EE[(A-\EE[A])(A^T - \EE[A^T])] \alpha_f]\|_2^2 =
      \|\Gamma \alpha_f\|_2^2 = \alpha_f^T \Gamma^T \Gamma \alpha_f,
\end{eqnarray*}
where $\Gamma = \Cov(A)$ is positive definite. This implies that $\alpha_f
\equiv 0$. Note that we have exploited 
above that $\mu_f = - \alpha_f^T \EE[A]$ since $\EE[(Y - f(X))] = 0$ for
all $f(\cdot)$ in $I$. Therefore,
we have that  
\begin{eqnarray*}
\EE[Y^a - f(X^a)] = \EE[Y - f(X)|\mathrm{do}(A = a)] = \EE[Y - f(X)|A = a]
  \equiv \mu_f,
\end{eqnarray*}
and thus being constant w.r.t. $a$.\hfill$\Box$

\subsubsection{Some empirical results.} 
To investigate the nature of discrete anchor variables, we consider a model
similar to (M2) and sample sizes $n = 300$ and $n_{\mathrm{out}} =
2000$ as before. The 
structural equation model is as for model (M2), except that the anchor
variables and the structural equation for $X$ are as follows.
\begin{figure}[!h]
\begin{center}
\includegraphics[scale=0.7]{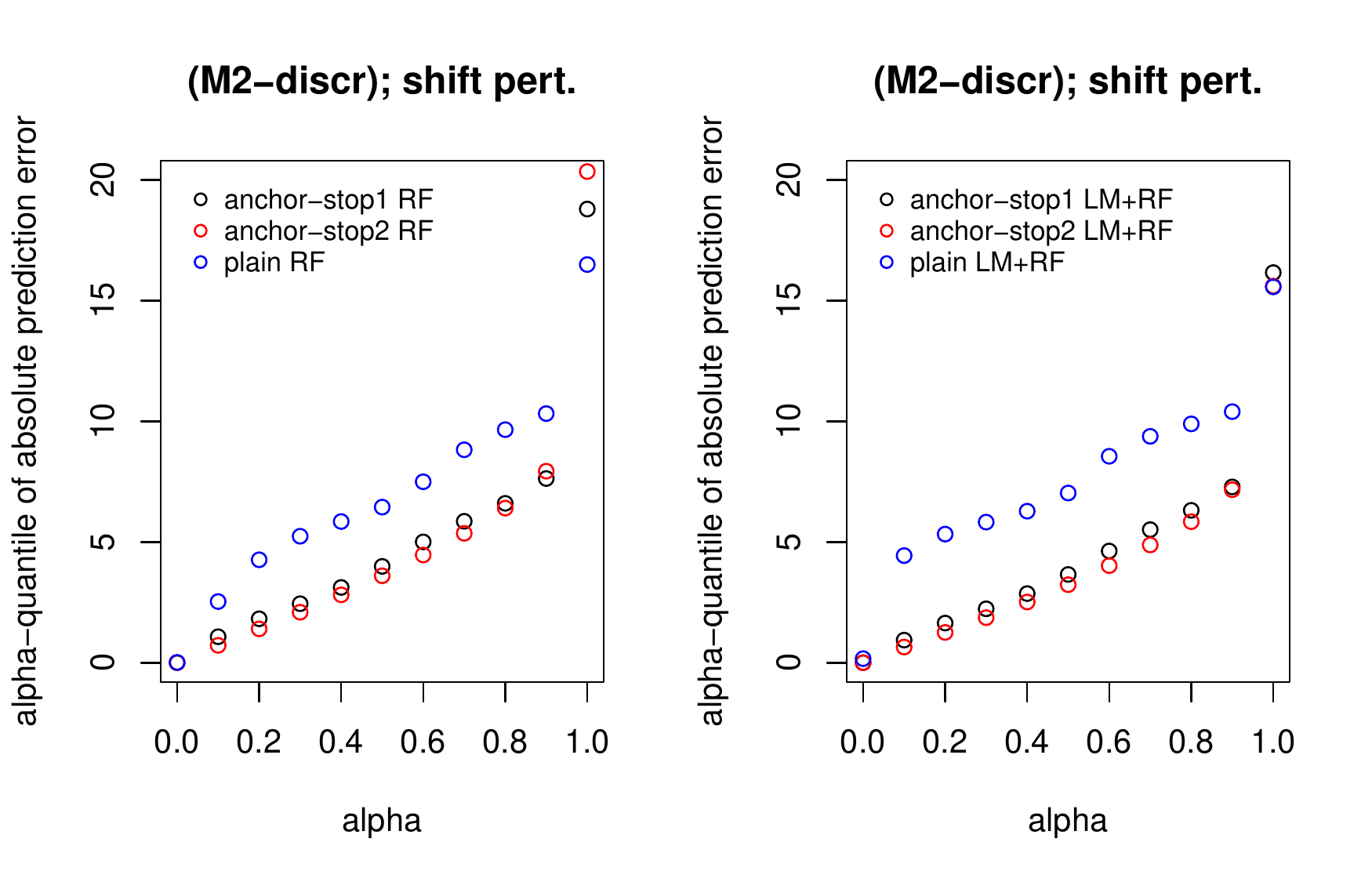}
\caption{Empirical $\alpha$-quantiles of $|Y_{\mathrm{out},i} -
  \hat{Y}_{\mathrm{out},i}|$ for $i = 1,\ldots, n_{\mathrm{out}} = 2000$,
  averaged over 100 independent simulation runs. Model (M2-discr) with
  3-fold amplification for $A_{\mathrm{out}}$. The Anchor
  Boosting algorithm is always with $\gamma = 7$, with Random
Forests (left) and with Linear Model + Random Forests
(right), and with the two stopping criteria from \eqref{stop1} (stop1) and \eqref{stop2} (stop2).}\label{fig.nonlin3}  
\end{center}
\end{figure}
\begin{description}
\item[(M2-discr)]
\begin{eqnarray*}
& &A_1 = (1,1,\ldots ,1,0,0,\ldots ,0)^T\ \mbox{where the first 150 entries are
  all $1$ and then $0$},\\
& &A_2 = (0,0,\ldots ,0,1,1,\ldots ,1)^T\ \mbox{where the first 150 entries are
  all $0$ and then $1$}. 
\end{eqnarray*}
The structural equation for $X$ is
\begin{eqnarray*}
X_j^{(i)} = 2 A_2^{(i)} - 2A_2^{(i)} + 2H^{(i)} + \eps{X,j}^{(i)},\ \eps_X
  \sim {\cal N}_{10}(0,0.5^2\mathrm{I}),
\end{eqnarray*}
where $i = 1,\ldots ,n$. 
The out-of-sample values of $A_{\mathrm{out}}$ are three times amplified: 
\begin{eqnarray*}
& &A_1 = (3,3,\ldots ,3,0,0,\ldots ,0)^T\ \mbox{where the first 150 entries are
  all $3$ and then $0$},\\
& &A_2 = (0,0,\ldots ,0,3,3,\ldots ,3)^T\ \mbox{where the first 150 entries are
  all $0$ and then $3$}. 
\end{eqnarray*}
\end{description}
The results are displayed in Figure \ref{fig.nonlin3} and are consistent
with the earlier results in Figures \ref{fig.nonlin1}-\ref{fig.nonlin2}. 

\section{Turning around the viewpoint}\label{sec.turnview}

One can switch to an alternative view and, although perhaps
thought-provoking, \emph{define} a (diluted) form of causality via the
invariance assumption. 

\begin{figure}[!htb]
\begin{center}
  \includegraphics[scale=0.4]{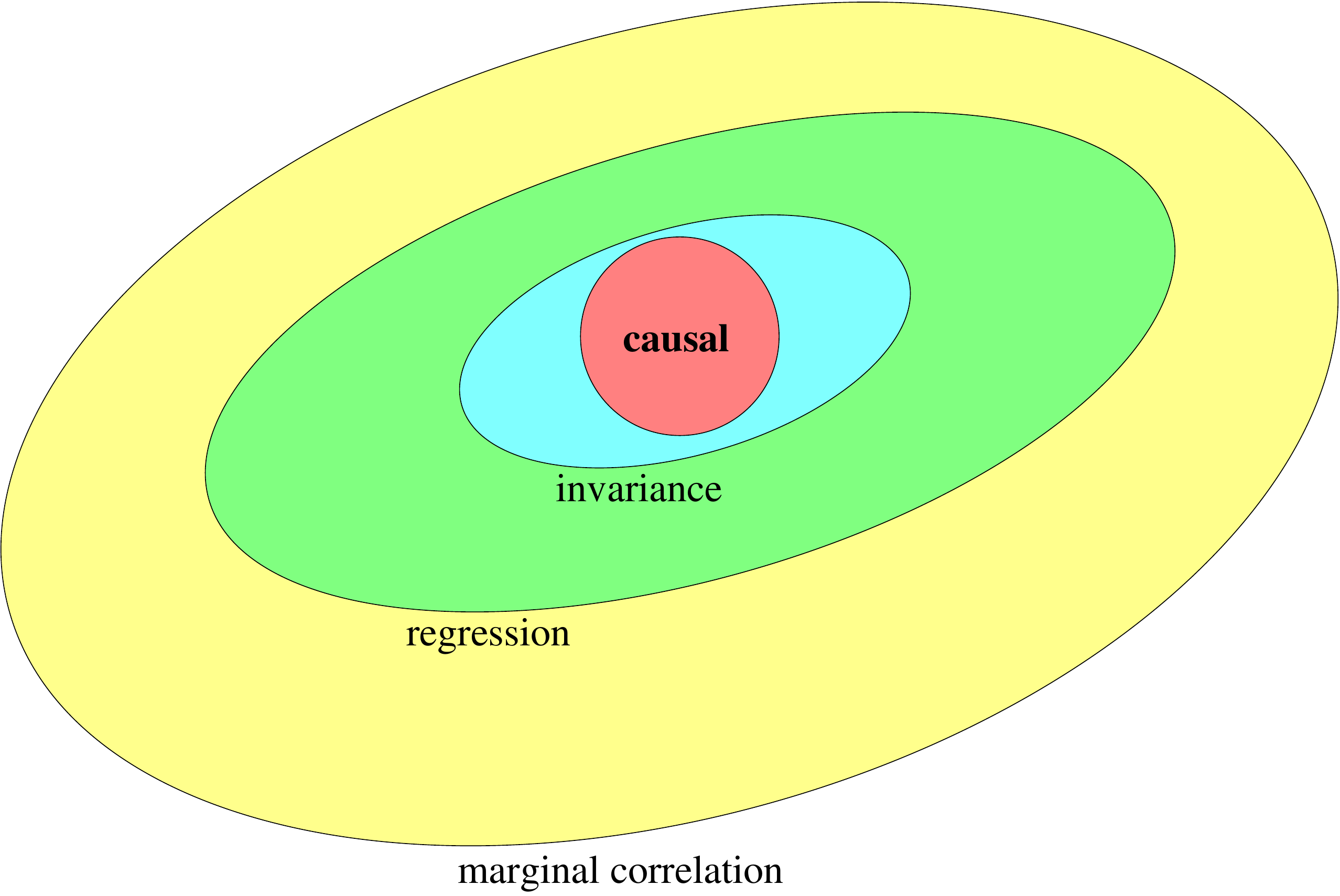}
  \caption{Illustration of various forms of associations between covariates
  $X$ and a response $Y$. Marginal correlation of components of $X$  with
  the response $Y$ is a weak notion; collecting
all variables with a non-zero regression coefficient is often
more informative as it measures the partial correlation of components of
$X$ with $Y$. Under a faithfulness assumption, the causal variables are a
subset of the relevant regression variables. In between is the notion of
invariance and the diluted form of causality: even when inferring causality is
impossible, obtaining the variables which satisfy invariance is often much
more useful in many applications.}\label{fig.screening}
\end{center}
\end{figure}

Consider a class of perturbations ${\cal F}$. Invariance (of variables) with
respect to the class of perturbations ${\cal F}$ is then defined as any set
of covariates such that the 
invariance assumption in Section \ref{sec.invariance} holds, that is 
\begin{eqnarray*}
{\cal L}(Y^e|X_S^e)\ \mbox{is the same for all}\ e \in {\cal F}.
\end{eqnarray*}
When ${\cal F}$ is sufficiently rich and fulfills the ad-hoc conditions 1 and 2
in Section \ref{sec.worstcaseoptim} or the assumption (B$({\cal F})$), then
the ${\cal F}$ invariance  
corresponds to the causality in the literature: this is just another
version of the worst case risk optimization viewpoint in
\eqref{causal-worstcaserisk}. If ${\cal F}$ is not sufficiently rich,
${\cal F}$-invariance does not coincide with the set of causal variables, 
see also Figure \ref{fig.screening}.

If ${\cal F}$ violates the ad-hoc conditions, then invariance as above does
not hold in general and 
is too demanding: but when restricting to
shift perturbations $v$ in an anchor regression model we still
obtain invariance of the residuals
\begin{eqnarray*}
  {\cal L}(Y^v - X^v \beta)\ \mbox{is the same for all}\ v \in {\cal F},
\end{eqnarray*}
where ${\cal F}$ is a subclass of shift
perturbations, see Proposition \ref{prop.shiftinv} and formula
\eqref{L2-uncor}.
We referred to this as the ${\cal F}$ diluted causality (see Section
\ref{subsec.diluted-causality}) but we emphasize here that it is a certain shift
invariance.  

Such invariances have interesting implications in terms
of interpretability and may lead to better insights in real
applications. Thus, even for cases where inferring causality is ill-posed and
non-identifiable, the invariance or diluted form of causality can provide 
more meaningful results and potentially contributes to an important aspect
of ``interpretable machine learning''. We illustrate this point also in
Figure \ref{fig.screening}. 

\section{Conclusions}

Causality can be phrased in terms of worst case prediction risk, see
Section \ref{subsec.predproblem}, showing that causal inference is linked
to a form of predictive robustness. The notion of (a certain) invariance
can be beneficially 
exploited for predictive robustness and hence also for causality. The
invariances themselves can be estimated from heterogeneous data: heterogeneity
is important and informative for inferring invariances. The paper includes
a review of recent results, points to the \texttt{R}-packages
\texttt{InvariantCausalPrediction} \citep{ICP-R}, \texttt{nonlinearICP}
\citep{nonlinICP-R} and
\texttt{seqICP} \citep{seqICP-R}, and contributes some new developments for nonlinear
problems in Section \ref{sec.nonlinanchor}.

Our contribution can be seen as dealing with ``statistics for perturbation (or
heterogeneous) data''. Even when causal inference is ill-posed, we show
here some attempts to obtain predictive robustness and more meaningful
approaches than what is provided by standard regression or curve fitting
technology.  

\section*{Acknowledgments}
We thank Nicolai Meinshausen, Jonas Peters, Dominik Rothenh\"ausler and
Niklas Pfister for many fruitful conversations.

\bibliographystyle{apalike}
\bibliography{references} 

\end{document}